\documentclass[12pt]{article}
\usepackage{jheppub}
\usepackage{ytableau}
\usepackage{comment}
\usepackage[vcentermath]{youngtab}
\usepackage{tikz}

\newcommand\be{\begin{equation}}
\newcommand\ee{\end{equation}}

\newcommand\Tr{\mathrm{Tr}}

\newcommand{\sfq}{q}
\newcommand{\sft}{\mathsf{t}}

\preprint{
RUP-26-6
}

\title{Quarter-indices for basic ortho-symplectic corners}
\abstract{
We study supersymmetric quarter-indices for corner configurations in 4d $\mathcal{N}=4$ super Yang-Mills theory with orthogonal and symplectic gauge groups. 
For the basic Y-junctions, we obtain exact closed-form expressions for the indices 
by making use of the Gustafson type integral formula and the Higgsing method. 
We demonstrate the equality of the quarter-indices between dual configurations, providing evidence for S-duality of the corner configurations. 
In the special fugacity limit, the indices admit an interpretation in terms of the vacuum characters 
of the W-algebras of type BCD, and the Lie superalgebra $\mathfrak{osp}(1|2N)$ as the corner vertex operator algebras. 
}
\author[a]{Yasuyuki Hatsuda}
\author[b]{and Tadashi Okazaki}
\emailAdd{yhatsuda@rikkyo.ac.jp, tokazaki@seu.edu.cn}
\affiliation[a]{Department of Physics, Rikkyo University, Toshima, Tokyo 171-8501, Japan}
\affiliation[b]{
School of Physics and Shing-Tung Yau Center, Southeast University,\\
Yifu Architecture Building, No.2 Sipailou, Xuanwu district, Nanjing, Jiangsu, 210096, China}
\begin{document}
\maketitle

\section{Introduction and summary}
\label{sec_intro}
Four-dimensional $\mathcal{N}=4$ super Yang-Mills (SYM) theory provides a natural framework 
for exploring non-perturbative physics in quantum field theory, most notably in connection with S-duality \cite{Montonen:1977sn,Osborn:1979tq} 
and the AdS/CFT correspondence \cite{Maldacena:1997re}. 
A natural class of such configurations arises from brane constructions involving D3-branes terminating on or intersecting $(p,q)$ 5-branes. 
These setups give rise to supersymmetric boundary conditions and interfaces in the 4d supersymmetric gauge theory \cite{Hanany:1996ie,Gaiotto:2008sa,Gaiotto:2008sd,Gaiotto:2008ak}. 
When multiple such interfaces intersect, one is led to consider corner configurations, 
in which distinct boundary conditions meet along codimension-two loci \cite{Gaiotto:2017euk,Creutzig:2017uxh,Prochazka:2017qum,Hanany:2018hlz,Gaiotto:2019jvo,Okazaki:2019bok}. 
These configurations can preserve a quarter of the original supersymmetry 
and provide a fertile ground for exploring the interplay between higher-dimensional gauge dynamics and lower-dimensional quantum field theories localized at the junction. 

From the field theoretic perspective, 
corner configurations define non-trivial sets of boundary conditions and junction data. 
Such systems can support protected sectors whose structure encodes intricate information about the parent four-dimensional theory. 
For configurations preserving two-dimensional supersymmetry along the junction, 
the observables capturing the structure of the corresponding BPS sectors are expected to be protected 
so that they furnish robust quantities from which one can extract information relevant to dualities and holographic aspects involving corner configurations. 
A particularly useful quantity in this context is the \textit{quarter-index} $\mathbb{IV}$, introduced in \cite{Gaiotto:2019jvo}. 
This supersymmetric index is defined for 4d $\mathcal{N}=4$ SYM theory in the quarter-BPS corner configuration 
in such a way that it generalizes the half-index \cite{Dimofte:2011py,Gang:2012yr} associated with a single boundary, 
and provides a systematic way to probe more general configurations involving multiple intersecting boundaries preserving $\mathcal{N}=(0,4)$ supersymmetry. 
Physically, the quarter-index involves contributions from the degrees of freedom localized at the corner, 
including both bulk and boundary fields subject to boundary conditions and additional junction degrees of freedom. 
As discussed in \cite{Gaiotto:2019jvo}, in appropriate fugacity limits, 
the quarter-indices admit an interpretation in terms of characters of the chiral algebras, the vertex operator algebras (VOAs) appearing at the corner 
\cite{Gaiotto:2017euk,Creutzig:2017uxh,Prochazka:2017qum}. 

In this work, we study the quarter-indices associated with corner configurations in 4d $\mathcal{N}=4$ SYM theory with orthogonal and symplectic gauge groups. 
We focus on Y-junctions realized by configurations of NS5-, D5-, and $(1,1)$ 5-branes, which divide space into three regions that may be occupied by D3-branes. 
Restricting to the simplest class of configurations, we consider setups in which a stack of multiple D3-branes occupies a single region. 
When D3-branes are suspended between the NS5-brane and the $(1,1)$ 5-brane, the gauge group is preserved. 
In this case, the Chern-Simons coupling induced by the $(1,1)$ 5-brane leads to a boundary gauge anomaly,  
which must be canceled by additional degrees of freedom localized at the corner. 
This cancellation can be achieved either by a 2d charged Fermi multiplet or 
by imposing the $\mathcal{N}=(0,4)$ Dirichlet boundary conditions on a 3d charged twisted hypermultiplet. 
The corresponding quarter-index is expressed as a non-trivial matrix integral, 
which we evaluate using the Gustafson type integral formula \cite{MR1139492,MR1199128}. 
On the other hand, when D3-branes occupy a region bounded by D5-branes, the gauge group is broken. 
In this situation, we compute the quarter-indices by employing the Higgsing procedure \cite{Gaiotto:2012xa}, 
as in the case of $U(N)$ gauge theories \cite{Gaiotto:2019jvo}. 
These corner configurations are expected to be related by S-duality, 
as follows from the Type IIB brane construction, 
in which S-duality acts as the $SL(2,\mathbb{Z})$ transformation on $(p,q)$ 5-branes 
and maps the induced boundary conditions in $\mathcal{N}=4$ SYM theory, 
while the O3-plane transforms non-trivially under the action of $SL(2,\mathbb{Z})$, 
leading to a map between orthogonal and symplectic gauge groups consistent with Langlands duality. 
We provide evidence for this expectation by explicitly demonstrating the agreement of the quarter-indices for pairs of dual configurations. 
Furthermore, we analyze these quarter-indices in suitable fugacity limits, 
in which they simplify and admit an interpretation in terms of vacuum characters of the vertex operator algebras associated with the corner. 
In particular, we find that they reproduce the expressions for the vacuum characters 
of the W-algebras associated with Lie algebras of type B, C and D, as well as the W-algebras associated with the Lie superalgebra $\mathfrak{osp}(1|2N)$.
This provides further evidence for the existence of the protected corner VOA sectors supported at the junctions of $\mathcal{N}=4$ SYM theories, 
as discussed by Gaiotto and Rap\v{c}\'{a}k \cite{Gaiotto:2017euk}.

\subsection{Structure}
This paper is organized as follows. 
In section \ref{sec_corners} we review the brane configurations of the ortho-symplectic Y-junction in Type IIB string theory 
and the associated gauge theory configurations, and provide the necessary background, including the quarter-indices. 
In sections \ref{sec_D}, \ref{sec_B}, \ref{sec_C} and \ref{sec_C2},  
we compute the quarter-indices for the basic ortho-symplectic Y-junctions associated with the Y-algebras  
$Y_{N,0,0}^+$, $\tilde{Y}_{N,0,0}^+$, $Y_{N,0,0}^-$, and $\tilde{Y}_{N,0,0}^-$ respectively. 
We present the exact closed-form formulae for the quarter-indices 
and demonstrate non-trivial agreement among these quarter-indices as strong evidence for dualities of the ortho-symplectic Y-junctions.
In the H-twist limit, it is shown that these quarter-indices precisely reproduce the vacuum characters of the corner VOAs. 

\subsection{Future directions}

\begin{itemize}

\item 
In this work, we have focused on the quarter-indices of basic ortho-symplectic Y-junctions 
in which the D3-branes occupy a single region and for which the Higgsing procedure can be applied. 
A natural and interesting direction for future work is to extend this analysis to more general Y-junctions, 
and further to more general corner setups, as discussed in \cite{Chung:2016pgt,Hanany:2018hlz,Gaiotto:2019jvo,Okazaki:2019bok} for unitary gauge theories. 
For such generic configurations, the gauge group remains unbroken in every duality frame, so there is no guarantee that the quarter-index admits an infinite product representation. Consequently, obtaining an exact closed-form expression becomes challenging. Nevertheless, the corresponding matrix integrals can still be evaluated, making this an interesting direction for future investigation. 

\item 
The lift of corner configurations to M-theory is expected to have interesting applications. 
In particular, the lift of the Y-junctions in $\mathcal{N}=4$ SYM theories with unitary gauge groups 
to M-theory has been discussed in \cite{Gaiotto:2019wcc,Gaiotto:2020dsq,Ishtiaque:2024orn}. 
From this viewpoint, the quarter-index for the basic corner of $\mathcal{N}=4$ $U(N)$ SYM theory 
can be interpreted as the special fugacity (twisted) limit \cite{Hayashi:2024aaf} of the superconformal index of 6d $\mathcal{N}=(2,0)$ theory of type A. 
Furthermore, the H-twist limit of the quarter-index coincides with the unrefined limit \cite{Kim:2012ava,Beem:2014kka} of the superconformal index, 
and is identified with the vacuum character of the $\mathcal{W}_{\mathfrak{gl}(N)}$ algebra, 
the W-algebra associated with $\mathfrak{gl}(N)$.\footnote{Up to the contribution of a decoupled free tensor multiplet associated with the center-of-mass degrees of freedom, this agrees with the vacuum character of the W-algebra of type A.} 
Upon lifting the configurations to M-theory, the presence of the O3-plane will give rise to a non-trivial background. 
It would be interesting to study the details of the M-theory lift of the ortho-symplectic Y-junctions. 

\item 
The inclusion of line operators in the present configurations leads to further interesting setups. 
The dualities of corner configurations are expected to admit an extension to configurations with line operators. 
From the viewpoint of the corner VOAs \cite{Gaiotto:2017euk}, line operators realize degenerate modules. 
In the M-theory lift, they map to M2-branes, which provide an important framework for testing the twisted holography \cite{Costello:2018zrm} 
via twisted M-theory \cite{Costello:2016nkh,Costello:2017fbo,Gaiotto:2019wcc}. 
A detailed analysis will be presented in forthcoming work.

\end{itemize}

\section{Ortho-symplectic corners}
\label{sec_corners}

\subsection{Brane setup}
\label{sec_brane}
In this section we describe the brane configurations relevant to our analysis of corner configurations of 4d $\mathcal{N}=4$ SYM theories 
and review their relevant properties that will be used throughout.

\subsubsection{O3-planes}
\label{sec_o3}
In order to realize the configurations of $\mathcal{N}=4$ SYM theories 
with orthogonal and symplectic gauge groups, we introduce orientifold planes. 
4d $\mathcal{N}=4$ SYM theories with orthogonal and symplectic gauge groups admit a realization in Type IIB string theory 
by placing D3-branes in the presence of an O3-plane \cite{Witten:1998xy,Feng:2000eq}. 
Such backgrounds are characterized by a pair of $\mathbb{Z}_2$-valued discrete fluxes, 
which may be identified with the discrete torsion parameters $\theta_{RR}$ and $\theta_{NS}$ associated to the RR and NSNS two-forms. 
The four resulting orientifold variants are related by the $SL(2,\mathbb{Z})$ duality group of Type IIB string theory. 

With non-trivial RR flux, one obtains an $\widetilde{\textrm{O3}}^-$-plane carrying $1/4$ unit of D3-brane charge.
A stack of $N$ D3-branes in this background realizes an $SO(2N+1)$ gauge theory.
Under $SL(2,\mathbb{Z})$, this orientifold is invariant under the $T$ transformation but is mapped by the $S$ transformation to an O3$^+$-plane with NS flux.

An O3$^+$-plane carries D3-brane charge $1/4$.
The low-energy theory on $N$ D3-branes placed in its background is a $USp(2N)$ gauge theory.
Consequently, $\mathcal{N}=4$ SYM theory with gauge group $SO(2N+1)$ is mapped under S-duality to $\mathcal{N}=4$ SYM theory with gauge group $USp(2N)$. 

Under the $SL(2,\mathbb{Z})$ transformation generated by $T$, the O3$^+$-plane is mapped to the $\widetilde{\textrm{O3}}^+$-plane, 
which carries a fractional D3-brane charge $1/4$.
The worldvolume theory on $N$ D3-branes in this background continues to be $\mathcal{N}=4$ $USp(2N)$ gauge theory, 
but with a non-trivial theta-angle, $\theta=\pi$, which we denote by $USp(2N)'$.
This orientifold configuration is invariant under the $S$ transformation. 

In the absence of discrete fluxes, the orientifold reduces to the O3$^-$-plane.
Unlike the other three variants, it carries a negative D3-brane charge $-1/4$.
The effective theory on $N$ D3-branes is $\mathcal{N}=4$ SYM with gauge group $O(2N)$ \cite{Garcia-Etxebarria:2015wns, Aharony:2016kai}.

The four possible types of O3-planes, 
together with the corresponding $\mathcal{N}=4$ SYM theories with orthogonal or symplectic gauge groups realized on D3-branes, 
are summarized as
\begin{align}
\label{O3_SYM}
\begin{array}{c|c|c|c|c}
&SO(2N+1)&USp(2N)&O(2N)&USp(2N)'\\ \hline 
\theta_{RR}&1/2&0&0&1/2 \\
\theta_{NS}&0&1/2&0&1/2 \\
\textrm{D3-brane charge}&1/4&1/4&-1/4&1/4 \\
\textrm{orientifold}&\widetilde{\textrm{O3}}^{-}&\textrm{O3}^+&\textrm{O3}^{-}&\widetilde{\textrm{O3}}^+ \\
\textrm{$S$ operation}&\textrm{O3}^+&\widetilde{\textrm{O3}}^{-}&\textrm{O3}^{-}&\widetilde{\textrm{O3}}^+ \\
\end{array}.
\end{align}

\subsubsection{Half 5-branes}
\label{sec_h5brane}
Now we proceed to determine the brane configuration explicitly.
We realize 4d $\mathcal{N}=4$ SYM theory of orthogonal or symplectic gauge group 
by considering a stack of D3-branes with a parallel O3-plane extending along the directions $0126$. 

Let us consider additional 5-branes; 
NS5-branes along the directions $012345$ and D5-branes along the directions $012789$. 
We take both to be localized at $x^6=0$ in such a way that 
the 5-brane gives rise to a boundary or an interface in 4d $\mathcal{N}=4$ SYM theory. 
The transverse directions decompose into $345$ and $789$, which keep the $SO(3)_C \times SO(3)_H$ subgroup of the $SO(6)_R$ R-symmetry. 
This symmetry is interpreted as the R-symmetry of 3d $\mathcal{N}=4$ supersymmetry. 
When the D3-branes end on a single 5-brane, one obtains half-BPS boundary conditions, 
whereas when they intersect it, the configuration gives rise to a half-BPS interface \cite{Gaiotto:2008sa,Gaiotto:2008sd,Gaiotto:2008ak}. 
More generally, when D3-branes intersect multiple 5-branes, one obtains a non-trivial 3d $\mathcal{N}=4$ gauge theory \cite{Hanany:1996ie}.

In the presence of an O3-plane, 5-branes in Type IIB string theory are subject to the orientifold projection, 
under which they effectively reduce to \textit{half 5-branes} localized at the orientifold fixed locus. 
The properties of the O3-planes, the half NS5- and D5-branes 
play a crucial role in determining the allowed half-BPS interfaces in 4d $\mathcal{N}=4$ SYM theory with orthogonal and symplectic gauge groups  \cite{Witten:1998xy,Gaiotto:2008sd,Gaiotto:2008ak}. 
In this setup, the half NS5- and D5-type interfaces act differently on the discrete data characterizing the orientifold background. 
The half NS5-brane induces a jump in the discrete NSNS flux, effectively exchanging the type of the O3-plane 
and thereby mapping orthogonal gauge groups to symplectic ones and vice versa. 
Consequently, the half NS5-type interfaces necessarily interpolate between orthogonal and symplectic  gauge theories.
In contrast, the half D5-branes modify the RR flux while preserving the orientifold type, 
and therefore implement boundary conditions or defects within a fixed class of gauge groups, 
such as $SO\leftrightarrow SO$ or $USp\leftrightarrow USp$,  possibly with rank shifts induced by half-D5 configurations. 
In summary, there exist four basic types of half-BPS interfaces for $\mathcal{N}=4$ SYM theories with gauge groups of type B, C, and D
\begin{enumerate}

\item $SO(2N+1)|USp(2M)'$ NS5-type

\item $O(2N)|USp(2M)$ NS5-type 

\item $USp(2N)|USp(2M)'$ D5-type 

\item $O(2N)|SO(2M+1)$ D5-type 

\end{enumerate}
The half NS5-type interface preserves the orthogonal and symplectic gauge groups on the two sides, 
and supports the half-hypermultiplet transforming in the bifundamental representation under these gauge groups. 
The half-D5-type interface preserves only one factor of the original product gauge group; 
in the case of unequal ranks, it preserves the gauge group of lower rank. 
When the ranks on the two sides coincide, the interface supports a (half-)hypermultiplet transforming in the fundamental representation. 
In contrast, when the ranks are different, no such (half-)hypermultiplet is present, and the interface instead involves a Nahm pole. 
Under S-duality of Type IIB string theory, D3-branes are invariant. 
However, taking into account the non-trivial transformation properties of O3-planes and half 5-branes, one is led to the following dualities among interfaces: 
\begin{align}
\textrm{$SO(2N+1)|USp(2M)'$ NS5-type} &\leftrightarrow \textrm{$USp(2N)|USp(2M)'$ D5-type}\\
\textrm{$SO(2N)|USp(2M)$ NS5-type} &\leftrightarrow \textrm{$SO(2N)|SO(2M+1)$ D5-type}. 
\end{align}
Strong evidence for these dualities has been reported in \cite{Hatsuda:2024lcc} from the computations of the half-indices.

\subsubsection{Ortho-symplectic Y-junctions}
\label{sec_Y}
We next consider an extension of the codimension-one interface configuration 
to a codimension-two corner configuration.
As discussed in \cite{Chung:2016pgt,Gaiotto:2017euk,Hanany:2018hlz,Gaiotto:2019jvo}, 
such a configuration can be realized by introducing NS5$'$-branes along the directions $016789$ and D5$'$-branes along the directions $0123456$, 
which preserves two-dimensional $\mathcal{N}=(0,4)$ chiral supersymmetry. 
All branes share the $01$ directions, defining the two-dimensional junction. 

Although a variety of corner configurations are in principle possible, in this work we restrict our attention to the Y-junction, 
which is a trivalent junction of the half NS5$'$-brane, half D5-brane, and half $(1,1)$ 5-brane, obtained by combining the half NS5$'$- and half D5-branes. 
The three faces of the resulting Y-shaped configuration, bounded respectively 
by the half NS5$'$- and $(1,1)$ 5-branes, the half $(1,1)$ 5- and D5-branes, and the half D5- and NS5$'$-branes, are filled with $L$, $M$, and $N$ D3-branes. 

\begin{figure}
\usetikzlibrary{shapes}
\scalebox{0.9}{
\begin{tikzpicture}
\filldraw [fill=yellow!20!white,draw=white] (-7.5,0) -- (-1.5,0) -- (-1.5,4) -- (-7.5,4);
\draw (-4.5,2)[dashed] -- (-4.5,4); 
\draw (-4.5,2)[dashed] -- (-7.5,2); 
\draw (-7.5,1.9) -- (-1.5,1.9); 
\node at (-4.5,1) {$SO(2L)$};
\node at (-6,3) {$USp(2M)'$};
\node at (-3,3) {$USp(2N)$};
\node at (-0.75,2) {NS5$'$};
\node at (-4.5,4.5) {D5};
\node at (-8.25,2) {$(1,1)$};
\filldraw [fill=yellow!20!white,draw=white] (1.5,0) -- (7.5,0) -- (7.5,4) -- (1.5,4);
\draw (4.5,2)[dashed] -- (4.5,4); 
\draw (4.5,2)[dashed] -- (1.5,2); 
\draw (1.5,1.9) -- (7.5,1.9); 
\node at (4.5,1) {$SO(2L+1)$};
\node at (3,3) {$USp(2M)$};
\node at (6,3) {$USp(2N)'$};
\node at (8.25,2) {NS5$'$};
\node at (4.5,4.5) {D5};
\node at (0.75,2) {$(1,1)$};
\filldraw [fill=yellow!20!white,draw=white] (-7.5,-7) -- (-1.5,-7) -- (-1.5,-3) -- (-7.5,-3);
\draw (-4.5,-5)[dashed] -- (-4.5,-3); 
\draw (-4.5,-5)[dashed] -- (-7.5,-5); 
\draw (-7.5,-5.1) -- (-1.5,-5.1); 
\node at (-4.5,-6) {$USp(2L)$};
\node at (-6,-4) {$SO(2M+1)$};
\node at (-3,-4) {$SO(2N)$};
\node at (-0.75,-5) {NS5$'$};
\node at (-4.5,-2.5) {D5};
\node at (-8.25,-5) {$(1,1)$};
\filldraw [fill=yellow!20!white,draw=white] (1.5,-7) -- (7.5,-7) -- (7.5,-3) -- (1.5,-3);
\draw (4.5,-5)[dashed] -- (4.5,-3); 
\draw (4.5,-5)[dashed] -- (1.5,-5); 
\draw (1.5,-5.1) -- (7.5,-5.1); 
\node at (4.5,-6) {$USp(2L)'$};
\node at (3,-4) {$SO(2M)$};
\node at (6,-4) {$SO(2N+1)$};
\node at (8.25,-5) {NS5$'$};
\node at (4.5,-2.5) {D5};
\node at (0.75,-5) {$(1,1)$};
\end{tikzpicture}
}
\caption{Four types of the ortho-symplectic Y-junctions. 
The configurations shown in the upper-left, upper-right, lower-left, and lower-right corners correspond to 
$Y_{L,M,N}^{+}$, $\tilde{Y}_{L,M,N}^{+}$, $Y_{L,M,N}^{-}$, and $\tilde{Y}_{L,M,N}^{-}$, respectively.}
\label{fig_Yjunction}
\end{figure}
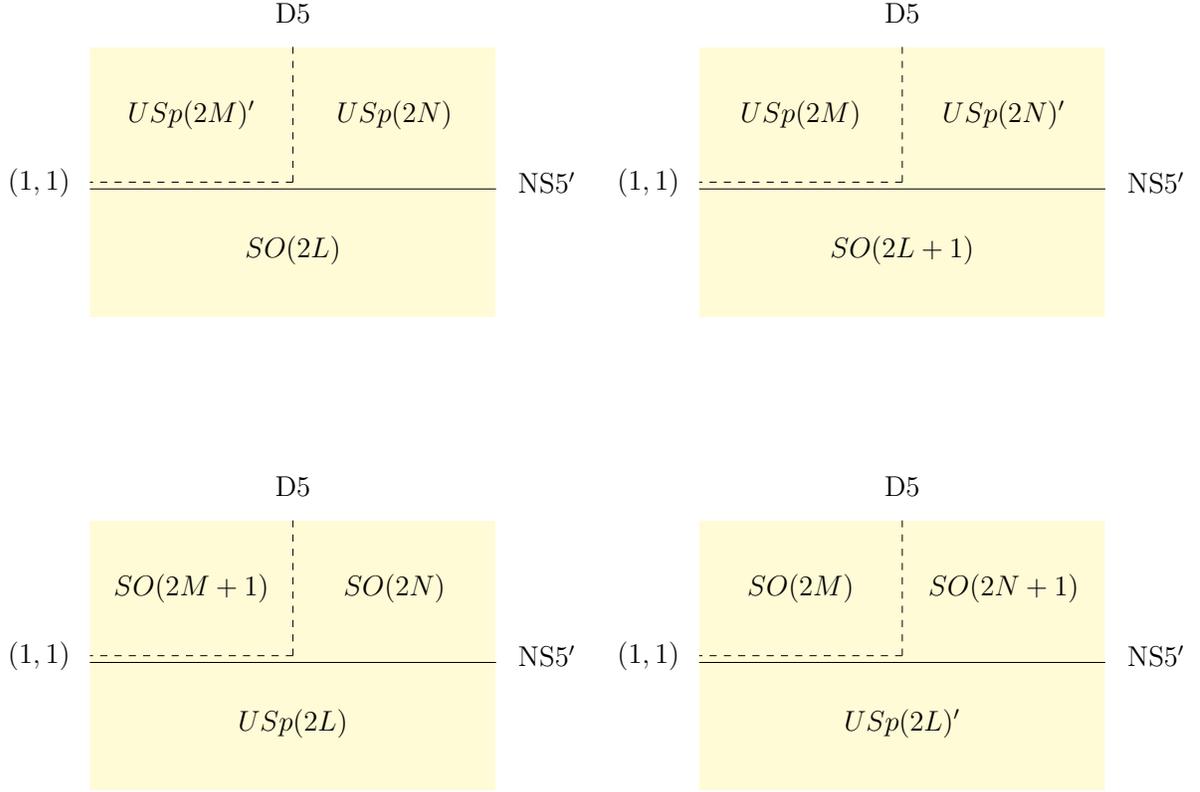

From the viewpoint of the low-energy effective theory on the D3-branes, 
this setup is described by a junction of three distinct half 5-brane interfaces, in 4d $\mathcal{N}=4$ gauge theories with orthogonal and symplectic gauge groups. 
While charge conservation in string theory requires that the half $(1,1)$ 5-brane be tilted in the $26$-plane \cite{Aharony:1997ju}, 
we will treat it schematically as extending along the $6$-direction, so that the configuration may be regarded as a T-shaped junction, as discussed in \cite{Gaiotto:2019jvo}. 
This description is appropriate insofar as the $(1,1)$ 5-brane induces boundary conditions analogous to those of an NS5$'$-type interface, 
differing only by a unit of boundary Chern-Simons coupling. 
There exist four types of Y-junctions, inheriting the structure of the four distinct interfaces (see Figure \ref{fig_Yjunction}).\footnote{The notation $SO(2N)$ in figures is used to indicate the underlying Lie algebra $\mathfrak{so}(2N)$. In the gauge theory analysis, we allow for both global forms $SO(2N)$ and $O(2N)$. } 
These Y-junctions lead to four classes of ortho-symplectic Y-algebras $Y_{L,M,N}^{\pm}$, $\tilde{Y}_{L,M,N}^{\pm}$ \cite{Gaiotto:2017euk} 
(see subsection \ref{sec_cornerVOA} for a review). 
Here, the superscripts $+$, $-$, and the presence or absence of a tilde refer to the type of O3-plane in the quadrant $x^6\ge0$ and $x^2\ge0$, 
where $N$ D3-branes are located. 
In other words, the configurations in the upper-left and upper-right corners of Figure \ref{fig_Yjunction} correspond to 
$Y_{L,M,N}^{+}$ and $\tilde{Y}_{L,M,N}^+$, respectively, 
while those in the lower-left and lower-right corners correspond to 
$Y_{L,M,N}^-$ and $\tilde{Y}_{L,M,N}^-$ respectively. 

A key feature of the Y-junction is the contribution of the Chern-Simons coupling induced by the $(1,1)$ 5-brane to the two-dimensional gauge anomaly localized at the junction. 
This contribution depends on the relative orientation of the boundary, namely on 
whether the corresponding boundary condition is supported in the region $x^2>0$ or $x^2<0$ \cite{Dimofte:2017tpi}.
More precisely, the Chern-Simons term induces an effective anomaly inflow that is equivalent to that of a two-dimensional chiral fermion. 
For the gauge symmetry associated with the region $x^2\ge0$, 
its contribution coincides with that of a left-moving (left-handed) chiral fermion, 
whereas for $x^2\le 0$ it matches that of a right-moving (right-handed) chiral fermion. 
In this way, the sign of the induced anomaly is determined by the orientation of the interface. 

Compared to the half-BPS interface, the Y-junction further contains the NS5$'$-brane located at $x^2=0$, 
which imposes the Neumann boundary conditions for the 3d $\mathcal{N}=4$ hypermultiplets. 
Besides, the D5-brane at $x^6=0$ imposes the Dirichlet-type boundary conditions for the 3d $\mathcal{N}=4$ twisted hypermultiplets. 
For the gauge group associated with the lower half-plane, 
the Dirichlet boundary conditions imposed on the 3d $\mathcal{N}=4$ twisted hypermultiplets 
contribute to the two-dimensional gauge anomaly at the junction. 
When the ranks of the gauge groups in the two upper quadrants coincide, 
the junction supports an additional Fermi multiplet transforming in the fundamental representation, whose contribution cancels the gauge anomaly in the lower half-plane. 

In the following, we study the ortho-symplectic Y-junctions, 
focusing on a basic class of corner configurations in which the orthogonal or symplectic gauge group in the lower half-plane is non-trivially preserved. 
We further consider the Y-junctions related by S-duality. 
In such dual configurations, the gauge group is completely broken, and the configuration involves a singular boundary condition of Nahm pole type.

\subsection{The corner VOA construction}
\label{sec_cornerVOA}
The correspondence between supersymmetric Y-junctions and vertex operator algebras (VOAs) was proposed by Gaiotto and Rap\v{c}\'{a}k \cite{Gaiotto:2017euk}. 
The starting point is a supersymmetric junction of 5-branes and D3-branes in Type IIB string theory, 
which engineers a supersymmetric corner of 4d $\mathcal N=4$ SYM theory. 
After performing the Geometric Langlands (GL) twist \cite{Kapustin:2006pk}, 
the protected local operators supported at the corner can form a  a two-dimensional VOA. 

An explicit realization of the corner VOA is obtained by describing the junction in terms of analytically continued Chern-Simons theories \cite{Mikhaylov:2014aoa}. 
Each boundary of the Chern-Simons theory supports an affine Kac-Moody algebra, 
while additional two-dimensional chiral degrees of freedom may be localized on the boundaries or interfaces separating different boundary conditions 
with potentially preserved gauge group. 
The corner VOA is then obtained by coupling these boundary and interface degrees of freedom and gauging the preserved gauge group through a BRST reduction 
or equivalently a coset construction. 
In this way, the VOA associated with a supersymmetric Y-junction generally can be realized as the BRST cohomology of a system 
consisting of affine current algebras, interface chiral fields, and ghost systems. 

This construction leads to a large family of VOAs, referred to as Y-algebras, 
which are labelled by the numbers of D3-branes occupying the three regions of the Y-junction. 
The ortho-symplectic corner VOAs admit a coset description of the form \cite{Gaiotto:2017euk}
\begin{align}
Y_{L,M,N}^+[\Psi]
&=\begin{cases}
\frac{W_{2N-2M}[OSp(2L|2N)_{-\Psi+2N-2L+2}]}{OSp(2L|2M)_{-\Psi+2M-2L+3}}&\textrm{$N>M$}\cr
\frac{W_{2M-2N}[OSp(2L|2M)_{-\Psi+2M-2L+1}]}{OSp(2L|2N)_{-\Psi+2M-2L+2}}&\textrm{$N<M$}\cr
\end{cases}, \\
\tilde{Y}_{L,M,N}^+[\Psi]
&=\begin{cases}
\frac{W_{2N-2M}[OSp(2L+1|2N)_{-\Psi+2N-2L+1}]}{OSp(2L+1|2N)_{-\Psi+2M-2L+2}}&\textrm{$N>M$}\cr
\frac{W_{2M-2N}[OSp(2L+1|2M)_{-\Psi+2M-2L}]}{OSp(2L+1|2N)_{-\Psi+2M-2L+1}}&\textrm{$N<M$}\cr
\end{cases}, \\
Y_{L,M,N}^-[\Psi]
&=\begin{cases}
\frac{W_{2N-2M-1}[OSp(2N|2L)_{\Psi-2N+2L+2}]}{OSp(2M+1|2L)_{\Psi-2M+2L}}&\textrm{$N>M+1$}\cr
\frac{W_{2M+1-2N}[OSp(2M+1|2L)_{\Psi-2M+2L}]}{OSp(2N|2L)_{\Psi-2N-2L+2}}&\textrm{$N<M$}\cr
\end{cases}, \\
\tilde{Y}_{L,M,N}^-[\Psi]&=Y_{L,N,M}^-[1-\Psi], 
\end{align}
where $W_\rho[G]$ denotes the quantum Drinfeld-Sokolov reduction associated with the Nahm pole of rank $\rho$. 
Here $\Psi$ is the GL parameter. 
The GL twist depends on a choice of parameter $t \in \mathbb{CP}^1$ characterizing the topological supercharge. 
Together with the complexified gauge coupling $\tau$, this defines an effective coupling parameter $\Psi$ 
which determines the levels of the affine algebras appearing in the corner VOA construction. 

Different brane configurations related by S-duality give rise to equivalent VOAs, leading to the triality property of the Y-algebras. 
Under S-duality, $\Psi$ transforms by fractional linear transformations, while the parameter $t$ is exchanged accordingly, 
leading to the triality structure of the Y-algebras.
Gaiotto and Rap\v{c}\'{a}k provided strong evidence for this correspondence 
by demonstrating the agreement of central charges, vacuum characters, and duality properties among the different realizations. 

In the case with $M=N=0$ we have the algebra $Y_{N,0,0}^*$, 
where the quantum Drinfeld-Sokolov reduction is absent since $W_0$ is trivial. 
This means that the algebras are realized via direct BRST quotients involving symplectic bosons. 
On the other hand, the corresponding brane configurations are mapped under the action of S-duality  
to configurations corresponding to the algebra $Y_{0,N,0}^*$.
In the latter case the denominators of the cosets frequently feature $OSp(0|0)$ (the trivial algebra) 
or $OSp(1|0) \simeq SO(1)$ (a single free fermion). 
Crucially, the quotient by $SO(1)$ is effectively trivial in this context. 
It merely decouples a free fermion without altering the chiral algebra. 
Consequently, the entire $Y_{0,N,0}^*$ family rigorously reduces to the W-algebras stripped of any residual coset complexities: 
\begin{align}
Y_{0,N,0}^+[\Psi]&=\frac{W_{2N}[OSp(0|2N)_{-\Psi+2N+1}]}
{OSp(0|0)_{-\Psi+2N+2}}\cong \mathcal{W}_{\mathfrak{usp}(2N)}, \\
\tilde{Y}_{0,N,0}^+[\Psi]&=\frac{W_{2N}[OSp(1|2N)_{-\Psi+2N}]}
{OSp(1|0)_{-\Psi+2N+1}}\cong \mathcal{W}_{\mathfrak{osp}(1|2N)}, \\ 
Y_{0,N,0}^-[\Psi]&=\frac{W_{2N}[OSp(2N+1|0)_{\Psi-2N}]}
{OSp(0|0)_{\Psi+2}}\cong \mathcal{W}_{\mathfrak{so}(2N+1)}, \\ 
\tilde{Y}_{0,N,0}^-[\Psi]&\cong \mathcal{W}_{\mathfrak{so}(2N)}. 
\end{align}

As discussed in \cite{Gaiotto:2019jvo}, 
the corresponding characters of the corner VOAs are expected to be obtained from the C-twist and H-twist limits of the supersymmetric quarter-indices. 
In the next subsection, we first review the definition and basic properties of the quarter-indices. 
We then show that the expected W-algebra characters are indeed reproduced from the quarter-indices.

\subsection{Quarter indices}
\label{sec_ind}

\subsubsection{Definition}
\label{sec_QindDef}
The main actors of this work are the quarter-indices $\mathbb{IV}$ that were introduced in \cite{Gaiotto:2019jvo}. 
They are the supersymmetric indices which count the BPS local operators 
residing on certain codimension-two defects in 4d ${\cal N}=4$ SYM theories, which preserve 2d $\mathcal{N}=(0,4)$ supersymmetry. 
The configurations involve two-dimensional junctions which lie at the intersection of multiple half-BPS interfaces or boundary conditions. 
This generalizes simpler indices such as the \textit{half-index} $\mathbb{II}$ of boundary or interface local operators \cite{Dimofte:2011py,Gang:2012yr,Gang:2012ff} 
and the \textit{full-index} $\mathbb{I}$ of bulk local operators \cite{Kinney:2005ej,Romelsberger:2005eg,Romelsberger:2007ec}. 
The latter can also be thought of as a specialization of the quarter-index to trivial junctions, possibly on a trivial interface. 

The definition of the quarter-index is \cite{Gaiotto:2019jvo} 
\begin{align}
\label{qind_def}
\mathbb{IV}(t,x;q)
&:={\Tr}_{\mathrm{Op}}(-1)^{F}q^{J+\frac{H+C}{4}}t^{H-C} x^{f}. 
\end{align}
Here the trace is taken over the cohomology of the chosen supercharges. $F$ is the fermion number, 
$J$ generates the $Spin(2)$ $\simeq$ $U(1)_{J}$ rotations in the two-dimensional plane on which the local operators are supported. $C$ and $H$ are the Cartan generators of the $SU(2)_{C}$ and $SU(2)_{H}$ R-symmetry groups. 
$f$ stands for a set of the Cartan generators of the flavor symmetry group. 
The choice of fugacities is such that, by the unitarity bound, 
the exponent of $q$ is strictly positive for any non-trivial local operator. 
Accordingly, the quarter-index is to be understood as a formal power series in $q$, 
whose coefficients are Laurent polynomials in the remaining fugacities. 

In the case where the 4d bulk theory is trivial and the 3d boundary theory is absent, 
the quarter-index reduces to the elliptic genus of 2d $\mathcal{N}=(0,4)$ supersymmetric field theory.
On the other hand, when the 4d bulk theory is trivial but a non-trivial 3d boundary theory is present, 
the quarter-index reduces to the half-index of $\mathcal{N}=(0,4)$ supersymmetric boundary condition for 3d $\mathcal{N}=4$ supersymmetric field theory.

To express the quarter-indices, we introduce the $q$-shifted factorial. 
We define 
\begin{align}
\label{qpoch_def}
(a;q)_{0}&:=1,\qquad
(a;q)_{n}:=\prod_{k=0}^{n-1}(1-aq^{k}),\qquad 
(q)_{n}:=\prod_{k=1}^{n}(1-q^{k}),
\nonumber \\
(a;q)_{\infty}&:=\prod_{k=0}^{\infty}(1-aq^{k}),\qquad 
(q)_{\infty}:=\prod_{k=1}^{\infty} (1-q^k), 
\end{align}
with $a, q \in \mathbb{C}$ and $|q|<1$. 
For simplicity we also use short-hand notations,
\begin{align}
(x^\pm; q)_\infty&=(x;q)_\infty (x^{-1};q)_\infty, \\
(x^\pm y^\pm;q)_\infty&=(xy;q)_\infty (xy^{-1};q)_\infty (x^{-1}y;q)_\infty  (x^{-1}y^{-1};q)_\infty.
\end{align}

\subsubsection{C-twist/H-twist limits}
\label{sec_CHtwist}
The $\mathcal{N}=(0,4)$ junctions admit deformations that are compatible with certain topological twists. 
In 4d $\mathcal{N}=4$ SYM theory with codimension-two defect preserving $\mathcal{N}=(0,4)$ supersymmetry, 
such twists are performed using the $SU(2)_C\times SU(2)_H$ R-symmetry. 
In particular, one may perform either the C-twist or the H-twist, 
obtained by identifying the rotation group along the defect with $SU(2)_C$ or $SU(2)_H$, respectively. 

After performing such a twist, one can select a scalar supercharge $\mathcal{Q}$ and restrict to its cohomology. 
A key feature of both the H-twist and the C-twist is that the generator of translations in the anti-holomorphic direction becomes $\mathcal{Q}$-exact. 
Consequently, correlation functions of operators in the $\mathcal{Q}$-cohomology are independent of 
the anti-holomorphic coordinate and depend only holomorphically on the position along the junction. 

This holomorphic dependence endows the space of local operators with the structure of a chiral algebra, 
or equivalently a vertex operator algebra (VOA), with the operator product expansion becoming meromorphic. 
In this way, the VOA arises naturally as the algebra of local operators in the $\mathcal{Q}$-cohomology after the twist. 

At the level of the supersymmetric index, the choice between the H-twist and the C-twist has a concrete manifestation. 
The supercharge used in the twist imposes a relation among the fugacities, leading to the specializations
\begin{align}
t &= q^{-\frac14} & \textrm{C-twist}, \\
t &= q^{\frac14} & \textrm{H-twist}, 
\end{align}
which are referred to as the \textit{C-twist limit} and \textit{H-twist limit} respectively \cite{Gaiotto:2019jvo}. 
These two specializations isolate precisely the protected subsectors captured by the two $\mathcal{Q}$-cohomologies associated with C- and H-twists.
As a consequence, the quarter-index reduces to the graded character of the corresponding VOA. 
In particular, this identification reproduces the characters of the corner VOAs introduced in \cite{Gaiotto:2017euk}, 
as well as those of the boundary VOAs constructed in \cite{Costello:2018fnz}.

\subsubsection{Examples}
As a concrete illustration, let us consider a corner configuration in the absence of orientifold planes. 
In this case, the relevant quarter-indices have been computed in \cite{Gaiotto:2019jvo,Okazaki:2019bok} for various junctions of half-BPS interfaces. 
These examples demonstrate how the general formalism of quarter-indices applies in practice, 
and provide explicit expressions for the protected spectrum of local operators residing at the corner. 

The simplest example is the Y-junction of $U(1)$ gauge theory, where no Nahm pole arises. 
For the $\left(\begin{smallmatrix}0&|&0\\ \hline &U(1)&\\ \end{smallmatrix}\right)$ Y-junction, 
the 4d $\mathcal{N}=4$ $U(1)$ gauge theory on a half-space $x^2\le 0$ should obey the Neumann boundary condition $\mathcal{N}'$ at $x^2=0$.\footnote{Following the conventions 
of \cite{Gaiotto:2019jvo}, we denote by $\mathcal{N}'$ the Neumann-type boundary condition at $x^2=0$ induced by an NS5$'$-brane, 
and by $\mathcal{D}$ the Dirichlet-type boundary condition at $x^6=0$ induced by a D5-brane. }
In the presence of the $(1,1)$ 5-brane in $x^6\le 0$, 
the boundary condition is shifted by a unit of Chern-Simons coupling, 
inducing a negative contribution to the $U(1)$ gauge anomaly in the region $x^2\le 0$. 
This anomaly is canceled by a charged Fermi multiplet localized at the junction. 
The corresponding quarter-index can then be evaluated as \cite{Gaiotto:2019jvo}
\begin{align}
\label{y100t}
\mathbb{IV}_{\mathcal{N}'\mathcal{D}}^{
\left(
\begin{smallmatrix}
0&|&0\\ \hline
&U(1)&\\
\end{smallmatrix}
\right)}(t;q)
&=
\frac{(q)_{\infty}}{(q^{\frac12} t^2;q)_{\infty}}
\oint \frac{ds}{2\pi is}
(q^{\frac12}s;q)_{\infty}(q^{\frac12}s^{-1};q)_{\infty}. 
\end{align}
On the other hand, for the $\left(\begin{smallmatrix}U(1)&|&0\\ \hline &0&\\ \end{smallmatrix}\right)$ Y-junction, 
a single D3-brane fills in the upper left quadrant of the plane. 
In this case, 4d $\mathcal{N}=4$ $U(1)$ gauge theory is defined on $x^2\ge 0$ with Neumann boundary condition. 
The D5-brane at $x^6=0$ imposes the Dirichlet boundary conditions, breaking the gauge group 
and projecting out the $SO(3)_C$ triplet scalars while preserving the $SO(3)_H$ triplet scalars. 
The resulting contributions to the quarter-index take the form \cite{Gaiotto:2019jvo}
\begin{align}
\label{y010t}
\mathbb{IV}_{\mathcal{N}'\mathcal{D}}^{
\left(
\begin{smallmatrix}
U(1)&|&0\\ \hline
&0&\\
\end{smallmatrix}
\right)}(t;q)
&=\frac{1}{(q^{\frac12}t^{2};q)_{\infty}}. 
\end{align}
The above two Y-junctions are related by the $S$-transformation 
of the $SL(2,\mathbb{Z})$ action of Type IIB string theory, 
and are thus expected to define dual corner configurations. 
Indeed, the corresponding quarter-indices (\ref{y100t}) and (\ref{y010t}) can be shown to coincide exactly (see \cite{Gaiotto:2019jvo} for the proof).
Moreover, in the H-twist limit, they reduce to 
\begin{align}
\mathbb{IV}_{\mathcal{N}'\mathcal{D}}^{
\left(
\begin{smallmatrix}
0&|&0\\ \hline
&U(1)&\\
\end{smallmatrix}
\right)}(t=q^{\frac14};q)
&=\mathbb{IV}_{\mathcal{N}'\mathcal{D}}^{
\left(
\begin{smallmatrix}
U(1)&|&0\\ \hline
&0&\\
\end{smallmatrix}
\right)}(t=q^{\frac14};q)
\nonumber\\
&=\frac{1}{(q)_{\infty}}=\chi_{\mathfrak{u}(1)}(q), 
\end{align}
which can be identified with the vacuum character of a single $U(1)$ current algebra \cite{Gaiotto:2017euk}. 

The analysis extends to $U(N)$ gauge theory, 
where the corresponding VOAs and quarter-indices have been investigated in \cite{Gaiotto:2017euk} and \cite{Gaiotto:2019jvo} respectively. 
The $\left(\begin{smallmatrix}0&|&0\\ \hline &U(N)&\\ \end{smallmatrix}\right)$ Y-junction 
consists of 4d $\mathcal{N}=4$ $U(N)$ gauge theory on $x^2\le 0$ with Neumann boundary condition $\mathcal{N}'$, 
along with the Fermi multiplet transforming as the fundamental representation that cancels the Chern-Simons induced gauge anomaly. 
The quarter-index takes the form \cite{Gaiotto:2019jvo}
\begin{align}
\label{yN00t}
&
\mathbb{IV}_{\mathcal{N}'\mathcal{D}}^{
\left(
\begin{smallmatrix}
0&|&0\\ \hline
&U(N)&\\
\end{smallmatrix}
\right)}(t;q)
\nonumber\\
&=
\frac{1}{N!}
\frac{(q)_{\infty}^N}{(q^{\frac12} t^2;q)_{\infty}^N}
\oint 
\prod_{i=1}^{N}\frac{ds_{i}}{2\pi is_{i}}
\prod_{i\neq j}
\frac{
\left(\frac{s_{i}}{s_{j}};q\right)_{\infty}
}
{
\left(q^{\frac12} t^2\frac{s_{i}}{s_{j}};q\right)_{\infty}
}
\prod_{i=1}^N
(q^{\frac12}s_{i};q)_{\infty}(q^{\frac12}s^{-1}_{i};q)_{\infty}. 
\end{align}
The matrix integral above contains the measure defining the Macdonald inner product, 
up to an additional factor $\prod_{i=1}^N (q^{\frac12}s_{i};q)_{\infty}(q^{\frac12}s^{-1}_{i};q)_{\infty}$. 
Upon removing this extra contribution, the integral reduces to the standard Macdonald-type integral \cite{MR1354144,MR1976581,MR1314036,MR1354956}, 
whose exact evaluation is immediate, and coincides with the half-index of the Neumann 
(or equivalently the regular Nahm pole) boundary condition for $U(N)$ SYM theory. 

Consider another type of the Y-junction $\left(\begin{smallmatrix}U(N)&|&0\\ \hline &0&\\ \end{smallmatrix}\right)$. 
As in the Abelian case, 4d $\mathcal{N}=4$ $U(N)$ gauge theory is defined on the half-space $x_2\ge 0$ with the Neumann boundary condition $\mathcal{N}'$. 
In this configuration, all $N$ D3-branes terminate on a single D5-brane at $x^6=0$, 
giving rise to a singular boundary condition associated with a regular Nahm pole of rank $N$, which completely breaks the gauge symmetry. 
In \cite{Gaiotto:2019jvo} it is proposed that 
the quarter-index of such singular corner configurations, including the Nahm pole, can be captured by the Higgsing method \cite{Gaiotto:2012xa}. 
Using this approach, the quarter-index for this junction is computed as \cite{Gaiotto:2019jvo}
\begin{align}
\label{y00Nt}
\mathbb{IV}_{\mathcal{N}'\mathcal{D}}^{
\left(
\begin{smallmatrix}
U(N)&|&0\\ \hline
&0&\\
\end{smallmatrix}
\right)}(t;q)
&=\prod_{k=1}^{N}\frac{1}{(q^{\frac{k}{2}}t^{2k};q)_{\infty}}. 
\end{align}
Since the brane configurations of the two Y-junctions are mapped into each other under the $SL(2,\mathbb{Z})$ transformation in Type IIB string theory, 
the corresponding corner configurations are expected to be dual corner configurations in $\mathcal{N}=4$ $U(N)$ SYM theory. 
This expectation is supported by the exact agreement of the corresponding quarter-indices (\ref{yN00t}) and (\ref{y00Nt}). 
To see it, we use the following integral formula (see Corollary 3.19 in \cite{MR3592530}):
\begin{align}
&\frac{1}{N!}
\frac{(\sfq)_\infty^N}{(\sft;\sfq)_\infty^N}
\oint 
\prod_{i=1}^{N}\frac{ds_{i}}{2\pi is_{i}}
\prod_{i\neq j}
\frac{
\left(\frac{s_{i}}{s_{j}};\sfq\right)_{\infty}
}
{
\left(\sft \frac{s_{i}}{s_{j}};\sfq\right)_{\infty}
}
\prod_{i=1}^N  \frac{(\sfq us_i^{-1};\sfq)_\infty (u^{-1} s_i;\sfq)_\infty}{(a s_i^{-1};\sfq)_\infty (b s_i;\sfq)_\infty}\notag \\
&=\prod_{k=1}^N \frac{(\sft^{k-1}au^{-1};\sfq)_\infty (\sfq \sft^{k-1}bu;\sfq)_\infty}{(\sft^k;\sfq)_\infty (\sft^{k-1}ab;\sfq)_\infty}.
\end{align}
By setting $\sft=q^{\frac{1}{2}}t^{2}$, $a=b=0$, $u=q^{-\frac{1}{2}}$ in this formula, we obtain the equality between (\ref{yN00t}) and (\ref{y00Nt}). 

In the H-twist limit, the quarter-indices (\ref{yN00t}) and (\ref{y00Nt}) reduce to the following expression:
\begin{align}
\mathbb{IV}_{\mathcal{N}'\mathcal{D}}^{
\left(
\begin{smallmatrix}
0&|&0\\ \hline
&U(N)&\\
\end{smallmatrix}
\right)}(t=q^{\frac14};q)
&=
\mathbb{IV}_{\mathcal{N}'\mathcal{D}}^{
\left(
\begin{smallmatrix}
U(N)&|&0\\ \hline
&0&\\
\end{smallmatrix}
\right)}(t=q^{\frac14};q)
\nonumber\\
&=\prod_{k=1}^{N}\frac{1}{(q^k;q)_{\infty}}=\chi_{\mathcal{W}_{\mathfrak{gl}(N)}}(q). 
\end{align}
This precisely coincides with the vacuum character of the W-algebra $\mathcal{W}_{\mathfrak{gl}(N)}$, 
that is $Y_{N,0,0}$ \cite{Gaiotto:2017euk}, associated with $\mathfrak{gl}(N)$. 

In the following sections, we generalize this class of quarter-index computations for the basic ortho-symplectic Y-junctions 
involving 4d $\mathcal{N}=4$ SYM theories with gauge groups of type B, C, and D. 
We show that the ortho-symplectic corners also admit elegant derivations of the quarter-indices via the Higgsing procedure.
The validity is established by rigorous proofs based on the Gustafson integral formula \cite{MR1139492} (also see \cite{MR1199128}):
\begin{align}
\label{Gustafson_int}
&G_{N}(a,b,c,d;\sfq,\sft)\notag \\
&:=\frac{1}{2^N N!}
\oint\prod_{i=1}^N \frac{ds_i}{2\pi i s_i}\frac{(s_i^{\pm 2};\sfq)_\infty}
{(a s_i^{\pm};\sfq)_\infty\,(b s_i^{\pm};\sfq)_\infty\,
(c s_i^{\pm};\sfq)_\infty\,(d s_i^{\pm};\sfq)_\infty}
\prod_{1\le i<j\le N}
\frac{(s_i^{\pm}s_j^{\pm};\sfq)_\infty}
{(\sft s_i^{\pm}s_j^{\pm};\sfq)_\infty} \notag\\
&=\frac{(\sft;\sfq)_\infty^N}{(\sfq;\sfq)_\infty^N}\prod_{j=1}^N
\frac{(\sft^{\,N+j-2}abcd;\sfq)_\infty}
{(\sft^j;\sfq)_\infty
(\sft^{j-1}ab ;\sfq)_\infty(\sft^{j-1}ac ;\sfq)_\infty(\sft^{j-1}ad ;\sfq)_\infty
(\sft^{j-1}bc ;\sfq)_\infty(\sft^{j-1}bd ;\sfq)_\infty(\sft^{j-1}cd ;\sfq)_\infty }.
\end{align}
We will see that suitable specializations of the parameters $(a,b,c,d)$ and $\sft$ reduce the matrix integrals appearing in the quarter-indices of the ortho-symplectic corners to closed product formulae. This fact allows us to show the rigorous agreement of the quarter-indices for the dual corner configurations.

\section{$Y_{N,0,0}^+$ and $\tilde{Y}_{0,N,0}^-$}
\label{sec_D}

\subsection{$Y_{N,0,0}^+$}

\begin{figure}
\usetikzlibrary{shapes}
\centering
\scalebox{0.9}{
\begin{tikzpicture}
\filldraw [fill=yellow!20!white,draw=white] (-3,0) -- (3,0) -- (3,2) -- (-3,2);
\draw (0,2)[dashed] -- (0,4); 
\draw (0,2)[dashed] -- (-3,2); 
\draw (-3,1.9) -- (3,1.9); 
\node at (0,1) {$SO(2N)$};
\node at (-1.5,3) {};
\node at (1.5,3) {};
\node at (4,2) {NS5$'$};
\node at (0,4.5) {D5};
\node at (-4,2) {$(1,1)$};
\end{tikzpicture}
}
\caption{The Y-junction for $Y_{N,0,0}^+$. }
\label{fig_Y+1}
\end{figure}
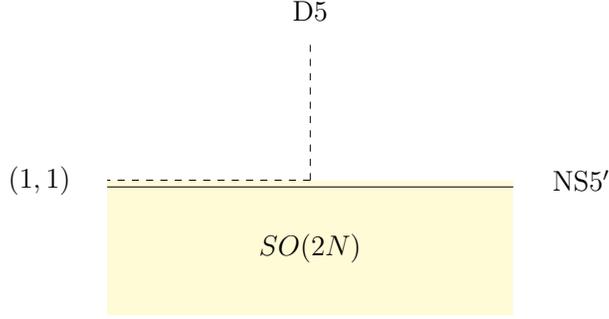

Consider the Y-junction for the ortho-symplectic Y-algebra $Y_{N,0,0}^+$. 
In the gauge theory description, 
this configuration involves a 4d $\mathcal{N}=4$ SYM theory with gauge group $O(2N)$ supported on the half-space $x^2\le 0$. 
The theory obeys the Neumann boundary condition $\mathcal{N}'$ at $x^2=0$. 
Also, there exists a Chern-Simons coupling, which leads to a gauge anomaly at the junction $x^2=x^6=0$. 
To ensure the gauge anomaly cancellation, the corner supports additional two-dimensional degrees of freedom, 
which can be described by the 2d Fermi multiplets transforming in the fundamental representation of the gauge group. 
We denote this Y-junction by $\left(
\begin{smallmatrix}
&0&|&0& \\ \hline 
&&SO(2N)&& \\
\end{smallmatrix}
\right)$. 
It is illustrated in Figure \ref{fig_Y+1}. 
%
%
%
%
%

The quarter-index of the Y-junction is evaluated as a matrix integral of the form 
\begin{align}
\label{qind_Y+}
&
\mathbb{IV}_{\mathcal{N}' \mathcal{D}}
^{\left(
\begin{smallmatrix}
&0&|&0& \\ \hline 
&&SO(2N)&& \\
\end{smallmatrix}
\right)}(t;q)
\nonumber\\
&=
\frac{1}{2^{N-1} N!}\frac{(q)_{\infty}^N}{(q^{\frac12}t^{2};q)_{\infty}^N}
\oint \prod_{i=1}^N \frac{ds_i}{2\pi is_i}
\prod_{i<j}
\frac{(s_i^{\pm}s_j^{\pm};q)_{\infty}}
{(q^{\frac12}t^{2}s_i^{\pm}s_j^{\pm};q)_{\infty}}
\prod_{i=1}^N (q^{\frac12}s_i^{\pm};q)_{\infty}. 
\end{align}
In particular, for $N=1$, it coincides with the quarter-index (\ref{y100t}) of the Y-junction for $U(1)$ gauge theory.

On the other hand, 
from the brane construction involving $N$ D3-branes in the background of an O3$^-$-plane 
one finds an $O(2N)$ gauge group as a $\mathbb{Z}_2$ gauging of the $SO(2N)$ gauge group 
\cite{Garcia-Etxebarria:2015wns, Aharony:2016kai}. 
The corresponding quarter-index for the Y-junction can be computed as follows:
\begin{align}
\mathbb{IV}_{\mathcal{N}' \mathcal{D}}
^{\left(
\begin{smallmatrix}
&0&|&0& \\ \hline 
&&O(2N)^{\pm}&& \\
\end{smallmatrix}
\right)}(t;q)
&=\frac12
\left(
\mathbb{IV}_{\mathcal{N}' \mathcal{D}}
^{\left(
\begin{smallmatrix}
&0&|&0& \\ \hline 
&&SO(2N)&& \\
\end{smallmatrix}
\right)}(t;q)
\pm 
\mathbb{IV}_{\mathcal{N}' \mathcal{D}}
^{\left(
\begin{smallmatrix}
&0&|&0& \\ \hline 
&&SO(2N)^-&& \\
\end{smallmatrix}
\right)}(t;q)
\right), 
\end{align}
where 
\begin{align}
\label{qind_Y+odd}
&
\mathbb{IV}_{\mathcal{N}' \mathcal{D}}
^{\left(
\begin{smallmatrix}
&0&|&0& \\ \hline 
&&SO(2N)^-&& \\
\end{smallmatrix}
\right)}(t;q)
\nonumber\\
&=\frac{1}{2^{N-1}(N-1)!}
\frac{(q)_{\infty}^{N-1}(-q;q)_{\infty}}
{(q^{\frac12}t^{2};q)_{\infty}^{N-1}(-q^{\frac12}t^{2};q)_{\infty}}
\oint \prod_{i=1}^{N-1}
\frac{ds_i}{2\pi is_i}
\frac{(s_i^{\pm};q)_{\infty} (-s_i^{\pm};q)_{\infty}}
{(q^{\frac12}t^{2}s_i^{\pm};q)_{\infty} (-q^{\frac12}t^{2}s_i^{\pm};q)_{\infty}}
\nonumber\\
&\times 
\prod_{1\le i<j\le N-1}
\frac{(s_i^{\pm}s_j^{\pm};q)_{\infty}}
{(q^{\frac12}t^2 s_i^{\pm}s_j^{\pm};q)_{\infty}}
(q^{\frac{1}{2}};q)_{\infty}(-q^{\frac{1}{2}};q)_{\infty} \prod_{i=1}^{N-1}(q^{\frac12}s_i^{\pm};q)_{\infty}. 
\end{align}
Note that we have $(-q;q)_{\infty}(q^{\frac{1}{2}};q)_{\infty}(-q^{\frac{1}{2}};q)_{\infty}=1$. 

\subsection{$\tilde{Y}_{0,N,0}^-$}

\begin{figure}
\usetikzlibrary{shapes}
\centering
\scalebox{0.9}{
\begin{tikzpicture}
\filldraw [fill=yellow!20!white,draw=white] (-3,2) -- (0,2) -- (0,4) -- (-3,4);
\draw (0,2)[dashed] -- (0,4); 
\draw (0,2)[dashed] -- (-3,2); 
\draw (-3,1.9) -- (3,1.9); 
\node at (0,1) {};
\node at (-1.5,3) {$SO(2N)$};
\node at (1.5,3) {};
\node at (4,2) {NS5$'$};
\node at (0,4.5) {D5};
\node at (-4,2) {$(1,1)$};
\end{tikzpicture}
}
\caption{The Y-junction for $\tilde{Y}_{0,N,0}^-$. }
\label{fig_tY-1}
\end{figure}
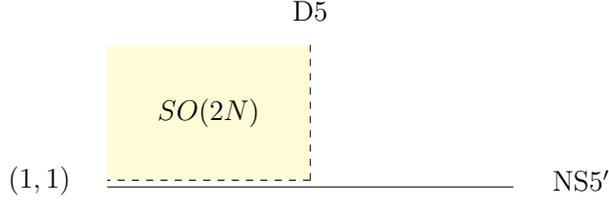

Next consider an alternative Y-junction configuration that realizes the $\tilde{Y}_{0,N,0}^{-}$ algebra 
in 4d $\mathcal{N}=4$ $SO(2N)$ SYM theory. 
In this configuration, the 4d $\mathcal{N}=4$ $SO(2N)$ SYM theory is supported on the half-space $x^2\ge 0$ and obeys Neumann boundary conditions at $x^2=0$. 
At $x^6=0$, the theory is subject to a singular boundary condition associated with a Nahm pole, which completely breaks the gauge symmetry. 
Hence there is no gauge group at the Y-junction. 
The configuration is shown in Figure \ref{fig_tY-1}. 
%
%
%
%
%

We first note that the Neumann-Dirichlet quarter-index for $\mathcal{N}=4$ $SO(2N)$ gauge theory takes the following form:
\begin{align}
\label{qind_SO2N}
\mathbb{IV}_{\mathcal{N}' \mathcal{D}}^{SO(2N)}(t,x_i;q)
&=\frac{1}{(q^{\frac12}t^2;q)_{\infty}^N}
\prod_{i<j}\frac{1}{(q^{\frac12}t^2x_i^{\pm}x_j^{\pm};q)_{\infty}}. 
\end{align}
This can be understood, as in the case of unitary gauge groups, by analyzing the degrees of freedom that survive the boundary condition. 
Although the corner configuration under consideration involves a singular boundary condition associated with a Nahm pole, 
it can be viewed as a deformation of the Dirichlet boundary condition. 
Thus we attempt to derive the quarter-index for this Y-junction 
by applying the Higgsing procedure to the Neumann-Dirichlet quarter-index and removing an appropriate set of 3d matter fields 
obeying the $\mathcal{N}=(0,4)$ Neumann boundary condition.  
By specializing the global fugacities as
\begin{align}
x_i&=q^{\frac{i-1}{2}}t^{2(i-1)}, 
\end{align}
the Neumann-Dirichlet quarter-index (\ref{qind_SO2N}) is factorized as
\begin{align}
&
\mathbb{IV}_{\mathcal{N}' \mathcal{D}}^{SO(2N)}(t,x_i=q^{\frac{i-1}{2}}t^{2(i-1)};q)
\nonumber\\
&=\mathbb{IV}_{\mathcal{N}' \mathcal{D}}
^{\left(
\begin{smallmatrix}
&SO(2N)&|&0& \\ \hline 
&&0&& \\
\end{smallmatrix}
\right)}(t;q)
\prod_{i=1}^{N-1}\mathbb{II}_N^{\textrm{3d HM}}(x_i=q^{\frac{2i-1}{4}} t^{2i-1})^{N-i}
\nonumber\\
&\times 
\prod_{i=1}^{2N-3}\mathbb{II}_{N}^{\textrm{3d HM}}(x_i=q^{\frac{2i-1}{4}}t^{2i-1})^{a_N(i)}. 
\end{align}
Here 
\begin{align}
\mathbb{II}_{N}^{\textrm{3d HM}}(x)
&=\frac{1}{(q^{\frac14}tx^{\pm};q)_{\infty}}
\end{align}
is the $\mathcal{N}=(0,4)$ Neumann half-index of the 3d hypermultiplet 
and 
\begin{align}
a_N(i) = \#\left\{\lambda \vdash i \;\middle|\; \lambda_1 \le N-2, \; \ell(\lambda)\le 2 \right\}
\end{align}
is the number of partitions of $i$ whose Young diagrams are contained in an $(N-2) \times 2$ rectangle. 
It is generated by the $q$-binomial coefficients \cite{MR1634067}
\begin{align}
\left(
\begin{matrix}
N\\
2\\
\end{matrix}
\right)_{q}&=\sum_i a_N(i)q^{i-1}. 
\end{align}
We remove the contribution of the 3d free hypermultiplets satisfying the Neumann boundary conditions 
by factoring out their half-indices. 
The remaining terms are then interpreted as the quarter-index associated with the Y-junction realizing the $ \tilde{Y}_{0,N,0}^- $ algebra. 
This procedure isolates the surviving degrees of freedom at the junction, leading to the following expression:
\begin{align}
\label{qind_tY-}
\mathbb{IV}_{\mathcal{N}' \mathcal{D}}
^{\left(
\begin{smallmatrix}
&SO(2N)&|&0& \\ \hline 
&&0&& \\
\end{smallmatrix}
\right)}(t;q)
&=
\frac{1}{(q^{\frac{N}{2}}t^{2N};q)_{\infty}}
\prod_{k=1}^{N-1} \frac{1}{(q^k t^{4k};q)_{\infty}}. 
\end{align}

To make the Higgsing procedure more transparent, we now present representative examples. 
Consider the case with $N=2$, corresponding to $SO(4)$. 
In this case the Neumann-Dirichlet quarter-index takes the form
\begin{align}
\label{qind_SO4}
\mathbb{IV}_{\mathcal{N}'\mathcal{D}}^{SO(4)}(t,x_i;q)
&=\frac{1}{(q^{\frac12}t^2;q)_{\infty}^2}
\frac{1}{(q^{\frac12}t^2x_1^{\pm}x_2^{\pm};q)_{\infty}}. 
\end{align}
We now implement the Higgsing specialization with $x_1=1$ and $x_2=q^{\frac12}t^2$. 
Then the Neumann-Dirichlet quarter-index (\ref{qind_SO4}) reduces to 
\begin{align}
\mathbb{IV}_{\mathcal{N}' \mathcal{D}}
^{\left(
\begin{smallmatrix}
&SO(4)&|&0& \\ \hline 
&&0&& \\
\end{smallmatrix}
\right)}(t;q)\times 
\mathbb{II}_{N}^{\textrm{3d HM}}(x=q^{\frac14}t)^2
\end{align}
By stripping off the $\mathcal{N}=(0,4)$ Neumann half-indices of the decoupled free hyper, 
we obtain the quarter-index
\begin{align}
\mathbb{IV}_{\mathcal{N}' \mathcal{D}}
^{\left(
\begin{smallmatrix}
&SO(4)&|&0& \\ \hline 
&&0&& \\
\end{smallmatrix}
\right)}(t;q)
&=\frac{1}{(qt^{4};q)_{\infty}^2}. 
\end{align}
As a slightly more involved example, we consider the case with $SO(6)$. 
We specialize the flavor fugacities as $x_1=1$, $x_2=q^{\frac12}t^2$, $x_3=qt^4$. 
Under this specialization, the corresponding Neumann-Dirichlet quarter-index admits the following factorization:
\begin{align}
\mathbb{IV}_{\mathcal{N}' \mathcal{D}}
^{\left(
\begin{smallmatrix}
&SO(6)&|&0& \\ \hline 
&&0&& \\
\end{smallmatrix}
\right)}(t;q)
\mathbb{II}_{N}^{\textrm{3d HM}}(x=q^{\frac14}t)^3
\mathbb{II}_{N}^{\textrm{3d HM}}(x=q^{\frac34}t^3)^2
\mathbb{II}_{N}^{\textrm{3d HM}}(x=q^{\frac54}t^5). 
\end{align}
We then find
\begin{align}
\mathbb{IV}_{\mathcal{N}' \mathcal{D}}
^{\left(
\begin{smallmatrix}
&SO(6)&|&0& \\ \hline 
&&0&& \\
\end{smallmatrix}
\right)}(t;q)&=\frac{1}{(qt^4;q)_{\infty}(q^{\frac32}t^6;q)_{\infty}(q^2t^8;q)_{\infty}}. 
\end{align}

This Y-junction $\left(
\begin{smallmatrix}
&0&|&0& \\ \hline 
&&SO(2N)&& \\
\end{smallmatrix}
\right)$ is mapped, through the $SL(2,\mathbb{Z})$ transformation in Type IIB string theory, to the configuration 
$\left(
\begin{smallmatrix}
&SO(2N)&|&0& \\ \hline 
&&0&& \\
\end{smallmatrix}
\right)$. 
Hence it is expected that the two configurations are dual configurations. 
Consistently, the quarter-index \eqref{qind_Y+} is expected to coincide with the quarter-index \eqref{qind_tY-},
\begin{align}
\label{eq:Y+=tY-}
\mathbb{IV}_{\mathcal{N}' \mathcal{D}}
^{\left(
\begin{smallmatrix}
&0&|&0& \\ \hline 
&&SO(2N)&& \\
\end{smallmatrix}
\right)}(t;q)
=
\mathbb{IV}_{\mathcal{N}' \mathcal{D}}
^{\left(
\begin{smallmatrix}
&SO(2N)&|&0& \\ \hline 
&&0&& \\
\end{smallmatrix}
\right)}(t;q)
\end{align}
We can directly show this equality by using the Gustafson integral formula \eqref{Gustafson_int}.
Let us consider the specialization
\begin{equation}
\begin{aligned}
(a,b,c,d)=(1,-1,-\sfq^{\frac{1}{2}}, 0),\qquad \sft=q^{\frac{1}{2}}t^2.
\end{aligned}
\end{equation}
Then the integrand in \eqref{Gustafson_int} becomes
\begin{align}
&\prod_{i=1}^N \frac{(s_i^{\pm 2};\sfq)_\infty}
{(s_i^{\pm};\sfq)_\infty\,(-s_i^{\pm};\sfq)_\infty\,
(-\sfq^{\frac{1}{2}}s_i^{\pm};\sfq)_\infty}
\prod_{1\le i<j\le N}\frac{(s_i^{\pm}s_j^{\pm};\sfq)_\infty}
{(q^{\frac{1}{2}}t^2 s_i^{\pm}s_j^{\pm};\sfq)_\infty}\notag \\
&=\prod_{i=1}^N (q^{\frac{1}{2}} s_i^{\pm};q)_\infty \prod_{1\le i<j\le N}\frac{(s_i^{\pm}s_j^{\pm};\sfq)_\infty}
{(q^{\frac{1}{2}}t^2 s_i^{\pm}s_j^{\pm};\sfq)_\infty},
\label{eq:G-integrand-1}
\end{align}
where we have used the identity:
\begin{align}
(x;q)_\infty (-x;q)_\infty (q^{\frac{1}{2}} x;q)_\infty (-q^{\frac{1}{2}} x;q)_\infty =(x^2; q)_\infty.
\label{eq:product-formula-1}
\end{align}
The result \eqref{eq:G-integrand-1} is nothing but the integrand of the quarter-index \eqref{qind_Y+}.
The Gustafson integral formula \eqref{Gustafson_int} is now given by
\begin{align}
&G_N(1,-1,-\sfq^{\frac{1}{2}}, 0;q, q^{\frac{1}{2}}t^2)\notag \\
&=\frac{1}{2^N N!}
\oint\prod_{i=1}^N \frac{ds_i}{2\pi i s_i}(q^{\frac{1}{2}} s_i^{\pm};q)_\infty \prod_{1\le i<j\le N}\frac{(s_i^{\pm}s_j^{\pm};\sfq)_\infty}
{(q^{\frac{1}{2}}t^2 s_i^{\pm}s_j^{\pm};\sfq)_\infty} \notag\\
&=\frac{(q^{\frac{1}{2}}t^2;q)_\infty^N}{(q;q)_\infty^N}\prod_{j=1}^N
\frac{1}
{(q^{\frac{j}{2}}t^{2j};q)_\infty
(-q^{\frac{j-1}{2}}t^{2j-2} ;q)_\infty(-q^{\frac{j}{2}}t^{2j-2} ;q)_\infty
(q^{\frac{j}{2}}t^{2j-2};q)_\infty}.
\end{align}
After some computations, one finds that the last product reduces to the simpler one
\begin{align}
&
\prod_{j=1}^N\frac{1}
{(q^{\frac{j}{2}}t^{2j};q)_\infty
(-q^{\frac{j-1}{2}}t^{2j-2} ;q)_\infty(-q^{\frac{j}{2}}t^{2j-2} ;q)_\infty
(q^{\frac{j}{2}}t^{2j-2};q)_\infty}\notag \\
&=
\frac{1}{2(q^{\frac{N}{2}} t^{2N};q)_\infty} \prod_{k=1}^{N-1} \frac{1}{(q^{k} t^{4k} ;q )_\infty}.
\end{align}
Therefore we finally obtain the equality \eqref{eq:Y+=tY-}.

In the H-twist limit, the quarter-index (\ref{qind_tY-}) becomes 
\begin{align}
\mathbb{IV}_{\mathcal{N}' \mathcal{D}}
^{\left(
\begin{smallmatrix}
&SO(2N)&|&0& \\ \hline 
&&0&& \\
\end{smallmatrix}
\right)}(t=q^{\frac14};q)
&=\frac{1}{(q^N;q)_{\infty}}\prod_{k=1}^{N-1}\frac{1}{(q^{2k};q)_{\infty}}
\nonumber\\
&=\chi_{\mathcal{W}_{\mathfrak{so}(2N)}}(q). 
\end{align}
This coincides with the vacuum character of the W-algebra $\mathcal{W}_{\mathfrak{so}(2N)}$ of type D \cite{Frenkel:1994em,MR1174415}, 
which is realized from the Neumann-Nahm junction as the ortho-symplectic Y-algebra $Y_{N,0,0}^+$ \cite{Gaiotto:2017euk} upon the quantum Drinfeld-Sokolov reduction. 

In an analogous manner, we can consider the quarter-index associated with a corner expected to arise from the brane configuration realizing $\mathcal{N}=4$ $O(2N)$ gauge theory. 
In this case, in addition to the expression (\ref{qind_tY-}) computed for the $SO(2N)$ junction, 
it is necessary to take into account the following Neumann-Dirichlet quarter-index:
\begin{align}
\mathbb{IV}_{\mathcal{N}' \mathcal{D}}^{O(2N)^{\pm}}(t,x_i;q)
&=\frac12\left(
\mathbb{IV}_{\mathcal{N}' \mathcal{D}}^{SO(2N)}(t,x_i;q)
\pm 
\mathbb{IV}_{\mathcal{N}' \mathcal{D}}^{SO(2N)^{-}}(t,x_i;q)
\right), 
\end{align}
where
\begin{align}
\label{qind_SO2N-}
&
\mathbb{IV}_{\mathcal{N}' \mathcal{D}}^{SO(2N)^{-}}(t,x_i;q)
\nonumber\\
&=\frac{1}{(q^{\frac12}t^2;q)_{\infty}^{N-1}(-q^{\frac12}t^2;q)_{\infty}}
\prod_{i=1}^{N}\frac{1}{(q^{\frac12}t^2x_i^{\pm};q)_{\infty} (-q^{\frac12}t^2 x_i^{\pm};q)_{\infty}}
\prod_{i<j}\frac{1}{(q^{\frac12}t^2x_i^{\pm}x_j^{\pm};q)_{\infty}}. 
\end{align}
Following the same procedure as before, 
we derive the quarter-index for this Y-junction by applying the Higgsing method to the Neumann-Dirichlet quarter-index (\ref{qind_SO2N-}) 
and stripping off the contributions of the 3d matter fields obeying the $\mathcal{N}=(0,4)$ Neumann boundary conditions. 
Here we specialize the global fugacities as
\begin{align}
x_i&=q^{\frac{i}{2}}t^{2i}. 
\end{align}
By again stripping off the contribution of the 3d hypermultiplets satisfying the Neumann boundary condition, 
we obtain the following expression: 
\begin{align}
\label{qind_tY-odd}
\mathbb{IV}_{\mathcal{N}' \mathcal{D}}
^{\left(
\begin{smallmatrix}
&SO(2N)^-&|&0& \\ \hline 
&&0&& \\
\end{smallmatrix}
\right)}(t;q)
&=
\frac{1}{(-q^{\frac{N}{2}}t^{2N};q)_{\infty}}
\prod_{k=1}^{N-1} \frac{1}{(q^k t^{4k};q)_{\infty}}. 
\end{align}

Again we expect from S-duality the equality between \eqref{qind_Y+odd} and \eqref{qind_tY-odd},
\begin{align}
\label{eq:Y+odd=tY-odd}
\mathbb{IV}_{\mathcal{N}' \mathcal{D}}
^{\left(
\begin{smallmatrix}
&0&|&0& \\ \hline 
&&SO(2N)^-&& \\
\end{smallmatrix}
\right)}(t;q)
=\mathbb{IV}_{\mathcal{N}' \mathcal{D}}
^{\left(
\begin{smallmatrix}
&SO(2N)^-&|&0& \\ \hline 
&&0&& \\
\end{smallmatrix}
\right)}(t;q)
\end{align}
To show it, we use the specialization
\begin{align}
(a,b,c,d)=(q^{\frac{1}{2}}t^{2},-q^{\frac{1}{2}}t^{2},-q^{\frac{1}{2}},0),\qquad \sft=q^{\frac{1}{2}}t^2.
\end{align}
in the Gustafson integral formula \eqref{Gustafson_int} for $N-1$. In this specialization, the integrand becomes
\begin{align}
&\prod_{i=1}^{N-1} \frac{(s_i^{\pm 2};\sfq)_\infty}
{(q^{\frac{1}{2}}t^{2} s_i^{\pm};\sfq)_\infty\,(-q^{\frac{1}{2}}t^{2} s_i^{\pm};\sfq)_\infty\,
(-q^{\frac{1}{2}}s_i^{\pm};\sfq)_\infty}
\prod_{1\le i<j\le N-1}\frac{(s_i^{\pm}s_j^{\pm};\sfq)_\infty}
{(q^{\frac{1}{2}}t^2 s_i^{\pm}s_j^{\pm};\sfq)_\infty}\notag \\
&=\prod_{i=1}^{N-1} \frac{(s_i^\pm;q)_\infty (-s_i^\pm;q)_\infty}{(q^{\frac{1}{2}}t^2 s_i^\pm;q)_\infty (-q^{\frac{1}{2}}t^2 s_i^\pm;q)_\infty}(q^{\frac{1}{2}}s_i^\pm;q)_\infty
\prod_{1\le i<j\le N-1}\frac{(s_i^{\pm}s_j^{\pm};\sfq)_\infty}
{(q^{\frac{1}{2}}t^2 s_i^{\pm}s_j^{\pm};\sfq)_\infty},
\label{eq:G-integrand-2}
\end{align}
where we have used the identity \eqref{eq:product-formula-1}. This is the same as the integrand of \eqref{qind_Y+odd}.
The Gustafson integral formula \eqref{Gustafson_int} is now given by
\begin{align}
&G_{N-1}(q^{\frac{1}{2}}t^{2},-q^{\frac{1}{2}}t^{2},-q^{\frac{1}{2}},0;q, q^{\frac{1}{2}}t^2)\notag \\
&=\frac{1}{2^{N-1} (N-1)!}
\oint\prod_{i=1}^{N-1} \frac{ds_i}{2\pi i s_i}\frac{(s_i^\pm;q)_\infty (-s_i^\pm;q)_\infty}{(q^{\frac{1}{2}}t^2 s_i^\pm;q)_\infty (-q^{\frac{1}{2}}t^2 s_i^\pm;q)_\infty}(q^{\frac{1}{2}}s_i^\pm;q)_\infty
\prod_{1\le i<j\le N-1}\frac{(s_i^{\pm}s_j^{\pm};\sfq)_\infty}
{(q^{\frac{1}{2}}t^2 s_i^{\pm}s_j^{\pm};\sfq)_\infty} \notag\\
&=\frac{(q^{\frac{1}{2}}t^2;q)_\infty^{N-1}}{(q;q)_\infty^{N-1}}
\prod_{j=1}^{N-1}\frac{1}
{(q^{\frac{j}{2}}t^{2j};q)_\infty
(-q^{\frac{j+1}{2}}t^{2j+2} ;q)_\infty(-q^{\frac{j+1}{2}}t^{2j} ;q)_\infty
(q^{\frac{j+1}{2}}t^{2j};q)_\infty}.
\end{align}
After some computations, one finds that the last product reduces to the simpler one
\begin{align}
&\prod_{j=1}^{N-1}\frac{1}
{(q^{\frac{j}{2}}t^{2j};q)_\infty
(-q^{\frac{j+1}{2}}t^{2j+2} ;q)_\infty(-q^{\frac{j+1}{2}}t^{2j} ;q)_\infty
(q^{\frac{j+1}{2}}t^{2j};q)_\infty}\notag \\
&=\frac{(-q^{\frac{1}{2}} t^2;q)_\infty}{(-q^{\frac{N}{2}} t^{2N};q)_\infty} \prod_{k=1}^{N-1} \frac{1}{(q^{k} t^{4k} ;q )_\infty}.
\end{align}
Therefore we finally obtain the equality \eqref{eq:Y+odd=tY-odd}.

\section{$\tilde{Y}_{N,0,0}^+$ and $\tilde{Y}_{0,N,0}^+$}
\label{sec_B}

\subsection{$\tilde{Y}_{N,0,0}^+$}
\begin{figure}
\usetikzlibrary{shapes}
\centering
\scalebox{0.9}{
\begin{tikzpicture}
\filldraw [fill=yellow!20!white,draw=white] (-3,0) -- (3,0) -- (3,2) -- (-3,2);
\draw (0,2)[dashed] -- (0,4); 
\draw (0,2)[dashed] -- (-3,2); 
\draw (-3,1.9) -- (3,1.9); 
\node at (0,1) {$SO(2N+1)$};
\node at (-1.5,3) {};
\node at (1.5,3) {};
\node at (4,2) {NS5$'$};
\node at (0,4.5) {D5};
\node at (-4,2) {$(1,1)$};
\end{tikzpicture}
}
\caption{The Y-junction for $\tilde{Y}_{N,0,0}^+$. }
\label{fig_tY+1}
\end{figure}
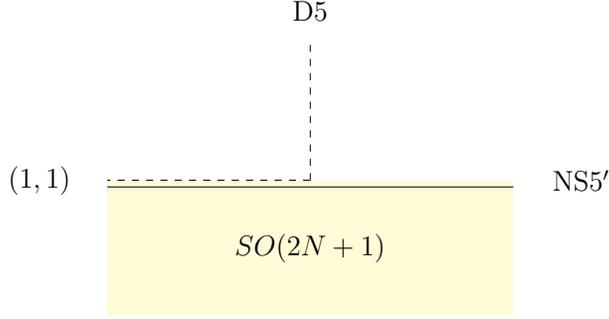

Now consider the Y-junction associated with the ortho-symplectic Y-algebra $\tilde{Y}_{N,0,0}^+$. 
In the Type IIB string theory configuration a stack of D3-branes is placed in the region $x^2\le 0$ in the background of an $\widetilde{\textrm{O3}}^{-}$-plane. 
The D3-branes terminate on two distinct 5-branes. 
At $x^2=0$ with $x^6\ge 0$, they end on a half NS5-brane, while along $x^2=x^6\le 0$, 
they end on a half $(1,1)$ 5-brane. 
Both 5-branes are compatible with the orientifold action and preserve a common $\mathcal{N}=(0,4)$ supersymmetry at the junction. 
In the gauge theory description, it is realized by a 4d $\mathcal{N}=4$ SYM theory with $SO(2N+1)$ gauge group on the half-space $x^2\le 0$, 
obeying the Neumann boundary condition $\mathcal{N}'$ at $x^2=0$.
The Chern-Simons coupling due to the half $(1,1)$ 5-brane induces a gauge anomaly localized at the junction $x^2=x^6=0$, 
which is canceled by a 2d Fermi multiplet in the fundamental representation of the $SO(2N+1)$ gauge group.
This Y-junction is denoted by $\left(
\begin{smallmatrix}
&0&|&0& \\ \hline 
&&SO(2N+1)&& \\
\end{smallmatrix}
\right)$, as shown in Figure \ref{fig_tY+1}. 

%
%
%
%
%

The quarter-index reads 
\begin{align}
\label{qind_tY+}
&
\mathbb{IV}_{\mathcal{N}'\mathcal{D}}
^{\left(
\begin{smallmatrix}
&0&|&0& \\ \hline 
&&SO(2N+1)&& \\
\end{smallmatrix}
\right)}(t;q)
\nonumber\\
&=
\frac{1}{2^N N!} \frac{(q)_{\infty}^N}{(q^{\frac12}t^{2};q)_{\infty}^N}
\oint 
\prod_{i=1}^N
\frac{ds_i}{2\pi is_i}
\frac{(s_i^{\pm};q)_{\infty}}
{(q^{\frac12}t^{2}s_i^{\pm};q)_{\infty}}
\prod_{i<j}
\frac{(s_i^{\pm}s_j^{\pm};q)_{\infty}}
{(q^{\frac12}t^{2}s_i^{\pm}s_j^{\pm};q)_{\infty}}
\nonumber\\
&\times (q^{\frac12};q)_{\infty}\prod_{i=1}^N
(q^{\frac12}s_i^{\pm};q)_{\infty}. 
\end{align}

\subsection{$\tilde{Y}_{0,N,0}^+$}
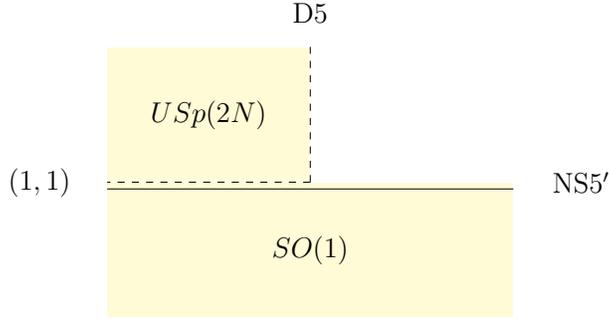
\begin{figure}
\usetikzlibrary{shapes}
\centering
\scalebox{0.9}{
\begin{tikzpicture}
\filldraw [fill=yellow!20!white,draw=white] (-3,2) -- (0,2) -- (0,4) -- (-3,4);
\filldraw [fill=yellow!20!white,draw=white] (-3,2) -- (3,2) -- (3,0) -- (-3,0);
\draw (0,2)[dashed] -- (0,4); 
\draw (0,2)[dashed] -- (-3,2); 
\draw (-3,1.9) -- (3,1.9); 
\node at (0,1) {$SO(1)$};
\node at (-1.5,3) {$USp(2N)$};
\node at (1.5,3) {};
\node at (4,2) {NS5$'$};
\node at (0,4.5) {D5};
\node at (-4,2) {$(1,1)$};
\end{tikzpicture}
}
\caption{The Y-junction for $\tilde{Y}_{0,N,0}^+$. }
\label{fig_tY+2}
\end{figure}

We consider the Y-junction realizing the $\tilde{Y}_{0,N,0}^+$ algebra. 
A Y-junction of the same type hosts $N$ D3-branes between a half D5-brane and a half $(1,1)$ 5-brane in the presence of an O3$^+$-plane. 
The $USp(2N)$ gauge theory lives in $x^2\ge 0$ with Neumann boundary conditions at $x^2=0$ 
and terminates at $x^6=0$ on a singular Nahm pole, 
completely breaking the gauge symmetry. 
Notably, even when no D3-branes are present between the half NS5- and half $(1,1)$ 5-branes, 
the corner configuration can nevertheless support non-trivial localized degrees of freedom.
The junction formally supports an $SO(1)$ gauge factor, which is trivial and contributes no local gauge degrees of freedom. 
Nevertheless, its presence indicates the existence of matter fields transforming under adjacent gauge groups, with $SO(1)$ acting trivially.
The Y-junction is illustrated in Figure \ref{fig_tY+2}. 

%
%
%
%
%

We derive the quarter-index of this Y-junction via a Higgsing procedure. 
As a starting point, we consider the associated Neumann-Dirichlet junction of $USp(2N)$ gauge theory. 
Although the $SO(1)$ gauge factor in $x^2\le 0$ is trivial, 
one finds an additional contribution from the $USp(2N)$ fundamental twisted hypermultiplet with Dirichlet boundary condition $D$. 
Accordingly, we take as our starting point the index of the following form: 
\begin{align}
\label{qind_USp2N_tHMD}
\mathbb{IV}_{\mathcal{N}' \mathcal{D}}^{USp(2N)}(t,x_i;q)
\times 
\prod_{i=1}^N
\mathbb{II}_{D}^{\textrm{3d tHM}}(x_i), 
\end{align}
where 
\begin{align}
&
\mathbb{IV}_{\mathcal{N}' \mathcal{D}}^{USp(2N)}(t,x_i;q)
=\frac{1}{(q^{\frac12}t^2;q)_{\infty}^N}
\prod_{i=1}^N \frac{1}{(q^{\frac12}t^2x_i^{\pm2};q)_{\infty}}
\prod_{i<j}
\frac{1}{(q^{\frac12}t^2x_i^{\pm}x_j^{\pm};q)_{\infty} }
\end{align}
is the quarter-index of the Neumann-Dirichlet junction of $USp(2N)$ gauge theory and 
\begin{align}
\mathbb{II}_{D}^{\textrm{3d tHM}}(x)
&=(q^{\frac34}tx^{\pm};q)_{\infty}
\end{align}
is the Dirichlet half-index of the twisted hypermultiplet. 
We implement the Higgsing by fixing the global fugacities as 
\begin{align}
x_i&=q^{\frac{2i-1}{4}}t^{2i-1}. 
\end{align}
Then the expression (\ref{qind_USp2N_tHMD}) factorizes into the following form: 
\begin{align}
&
\mathbb{IV}_{\mathcal{N}' \mathcal{D}}^{USp(2N)}(t,x_i=q^{\frac{2i-1}{4}}t^{2i-1};q)
\times 
\prod_{i=1}^N
\mathbb{II}_{D}^{\textrm{3d tHM}}(x_i=q^{\frac{2i-1}{4}}t^{2i-1})
\nonumber\\
&=\mathbb{IV}_{\mathcal{N}' \mathcal{D}}
^{\left(
\begin{smallmatrix}
&USp(2N)&|&0& \\ \hline 
&&SO(1)&& \\
\end{smallmatrix}
\right)}(t;q)
\prod_{i=1}^N \mathbb{II}_N^{\textrm{3d HM}}(x=q^{\frac{2i-1}{2}}t^{4i-2})
\nonumber\\
&\times 
\prod_{i=1}^{N-1} \mathbb{II}_N^{\textrm{3d HM}}(x=q^{\frac{2i-1}{4}}t^{2i-1})^{N-i}
\prod_{i=1}^{2N-3}\mathbb{II}_{N}^{\textrm{3d HM}}(x=q^{\frac{2i+1}{4}}t^{2i+1})^{a_N(i)}
\nonumber\\
&\times F_0(q) \prod_{i=1}^{N-1}F(x_i=q^{\frac{i}{2}}t^{2i}), 
\end{align}
where 
\begin{align}
F_0(q)=(q^{\frac12};q)_{\infty}
\end{align}
can be interpreted as a contribution from the neutral Fermi multiplet in the trivial representation, 
i.e. the zero-weight sector 
and 
\begin{align}
F(x)=(q^{\frac12}x^{\pm};q)_{\infty}
\end{align}
is the index of the 2d Fermi multiplet. 
After factoring out the decoupled contribution, 
we are led to the following expressions for the quarter-index of the $\tilde{Y}_{0,N,0}^+$ junction: 
\begin{align}
\label{qind_tY+2}
\mathbb{IV}_{\mathcal{N}' \mathcal{D}}
^{\left(
\begin{smallmatrix}
&USp(2N)&|&0& \\ \hline 
&&SO(1)&& \\
\end{smallmatrix}
\right)}(t;q)
&=(q^{\frac{N+1}{2}} t^{2N};q)_{\infty}
\prod_{k=1}^N
\frac{1}{(q^k t^{4k};q)_{\infty}}. 
\end{align}
We now observe that the Y-junction for the $\tilde{Y}_{0,N,0}^+$ algebra is S-dual to the Y-junction for the $\tilde{Y}_{N,0,0}^+$ algebra.  
In fact, the expression (\ref{qind_tY+2}) precisely matches the quarter-index (\ref{qind_tY+}), providing a non-trivial consistency check of our Higgsing calculation,
\begin{align}
\label{eq:tY+=tY+2}
\mathbb{IV}_{\mathcal{N}'\mathcal{D}}
^{\left(
\begin{smallmatrix}
&0&|&0& \\ \hline 
&&SO(2N+1)&& \\
\end{smallmatrix}
\right)}(t;q)
=
\mathbb{IV}_{\mathcal{N}' \mathcal{D}}
^{\left(
\begin{smallmatrix}
&USp(2N)&|&0& \\ \hline 
&&SO(1)&& \\
\end{smallmatrix}
\right)}(t;q)
\end{align}
Let us prove this equality by using the Gustafson integral formula \eqref{Gustafson_int}.
We consider the specialization
\begin{equation}
\begin{aligned}
(a,b,c,d)=(q^{\frac{1}{2}}t^{2},-1,-\sfq^{\frac{1}{2}}, 0),\qquad \sft=q^{\frac{1}{2}}t^2.
\end{aligned}
\end{equation}
Then the integrand in \eqref{Gustafson_int} becomes
\begin{align}
&\prod_{i=1}^N \frac{(s_i^{\pm 2};\sfq)_\infty}
{(q^{\frac{1}{2}}t^{2} s_i^{\pm};\sfq)_\infty\,(-s_i^{\pm};\sfq)_\infty\,
(-\sfq^{\frac{1}{2}}s_i^{\pm};\sfq)_\infty}
\prod_{1\le i<j\le N}\frac{(s_i^{\pm}s_j^{\pm};\sfq)_\infty}
{(q^{\frac{1}{2}}t^2 s_i^{\pm}s_j^{\pm};\sfq)_\infty}\notag \\
&=\prod_{i=1}^N  \frac{(s_i^{\pm};\sfq)_\infty (\sfq^{\frac{1}{2}}s_i^{\pm};\sfq)_\infty}
{(q^{\frac{1}{2}}t^{2} s_i^{\pm};\sfq)_\infty
} \prod_{1\le i<j\le N}\frac{(s_i^{\pm}s_j^{\pm};\sfq)_\infty}
{(q^{\frac{1}{2}}t^2 s_i^{\pm}s_j^{\pm};\sfq)_\infty}.
\label{eq:G-integrand-3}
\end{align}
This is the integrand of the quarter-index \eqref{qind_tY+}.
The Gustafson integral formula \eqref{Gustafson_int} is now given by
\begin{align}
&G_N(q^{\frac{1}{2}}t^{2},-1,-\sfq^{\frac{1}{2}}, 0;q, q^{\frac{1}{2}}t^2)\notag \\
&=\frac{1}{2^N N!}
\oint\prod_{i=1}^N \frac{ds_i}{2\pi i s_i}\frac{(s_i^{\pm};\sfq)_\infty (\sfq^{\frac{1}{2}}s_i^{\pm};\sfq)_\infty}
{(q^{\frac{1}{2}}t^{2} s_i^{\pm};\sfq)_\infty
} \prod_{1\le i<j\le N}\frac{(s_i^{\pm}s_j^{\pm};\sfq)_\infty}
{(q^{\frac{1}{2}}t^2 s_i^{\pm}s_j^{\pm};\sfq)_\infty} \notag\\
&=\frac{(q^{\frac{1}{2}}t^2;q)_\infty^N}{(q;q)_\infty^N}\prod_{j=1}^N
\frac{1}
{(q^{\frac{j}{2}}t^{2j};q)_\infty
(-q^{\frac{j}{2}}t^{2j} ;q)_\infty(-q^{\frac{j+1}{2}}t^{2j} ;q)_\infty
(q^{\frac{j}{2}}t^{2j-2};q)_\infty}.
\end{align}
After some computations, one finds that the last product reduces to the simpler one
\begin{align}
&
\prod_{j=1}^N
\frac{1}
{(q^{\frac{j}{2}}t^{2j};q)_\infty
(-q^{\frac{j}{2}}t^{2j} ;q)_\infty(-q^{\frac{j+1}{2}}t^{2j} ;q)_\infty
(q^{\frac{j}{2}}t^{2j-2};q)_\infty}\notag \\
&=
\frac{(q^{\frac{N+1}{2}}t^{2N};q)_\infty}{(q^{\frac{1}{2}};q)_\infty} \prod_{k=1}^{N} \frac{1}{(q^{k} t^{4k} ;q )_\infty}.
\end{align}
Therefore we finally obtain the equality \eqref{eq:tY+=tY+2}.

After implementing the H-twist, the quarter-index (\ref{qind_tY+2}) or equivalently (\ref{qind_tY+}) simplifies to
\begin{align}
\label{ch_tY+2}
\mathbb{IV}_{\mathcal{N}' \mathcal{D}}
^{\left(
\begin{smallmatrix}
&USp(2N)&|&0& \\ \hline 
&&SO(1)&& \\
\end{smallmatrix}
\right)}(t=q^{\frac14};q)
&=(q^{N+\frac{1}{2}};q)_{\infty}\prod_{k=1}^N \frac{1}{(q^{2k};q)_{\infty}}. 
\end{align}
Here we identify the factor $(q^{N+\frac12};q)_{\infty}$ 
with the fermionic contribution arising from the odd sector of the principal quantum Drinfeld-Sokolov (DS) reduction of $\mathfrak{osp}(1|2N)$. 
The odd part of $\mathfrak{osp}(1|2N)$ transforms in the fundamental representation of $USp(2N)$, which has dimension $2N$. 
Under the principal $\mathfrak{sl}(2)$ embedding, 
this representation becomes an irreducible representation of spin $j$ $=$ $\frac{2N-1}{2}$. 
The DS reduction maps this odd part to a single fermionic generator of conformal dimension $h$ $=$ $j+1$ $=$ $N+\frac12$. 
The corresponding Fock space contributes to the vacuum character as $(q^h;q)_{\infty}$, 
which gives rise to the observed factor. 
The remaining factor is the vacuum character of the W-algebra $\mathcal{W}_{\mathfrak{usp}(2N)}$ of type C \cite{Frenkel:1994em,MR1174415}. 

The resulting expression (\ref{ch_tY+2}) admits a natural interpretation as a super extension of the W-algebra. 
More precisely, it is consistent with the structure of the principal DS reduction of the Lie superalgebra $\mathfrak{osp}(1|2N)$. 
In particular, the bosonic sector reproduces the generators of $\mathcal{W}_{\mathfrak{usp}(2N)}$, 
while the odd part gives rise to an additional fermionic generator of conformal dimension $N+\frac12$ 
so that the Y-algebra $\tilde{Y}_{0,N,0}^+$ can be regarded 
as a realization of the super W-algebra $\mathcal{W}_{\mathfrak{osp}(1|2N)}$ \cite{Gaiotto:2017euk}.

\section{$Y_{N,0,0}^-$ and $Y_{0,N,0}^-$}
\label{sec_C}

\subsection{$Y_{N,0,0}^-$}

\begin{figure}
\usetikzlibrary{shapes}
\centering
\scalebox{0.9}{
\begin{tikzpicture}
\filldraw [fill=yellow!20!white,draw=white] (-3,0) -- (3,0) -- (3,2) -- (-3,2);
\filldraw [fill=yellow!20!white,draw=white] (-3,2) -- (0,2) -- (0,4) -- (-3,4);
\draw (0,2)[dashed] -- (0,4); 
\draw (0,2)[dashed] -- (-3,2); 
\draw (-3,1.9) -- (3,1.9); 
\node at (0,1) {$USp(2N)$};
\node at (-1.5,3) {$SO(1)$};
\node at (1.5,3) {};
\node at (4,2) {NS5$'$};
\node at (0,4.5) {D5};
\node at (-4,2) {$(1,1)$};
\end{tikzpicture}
}
\caption{The Y-junction for $Y_{N,0,0}^-$. }
\label{fig_Y-1}
\end{figure}
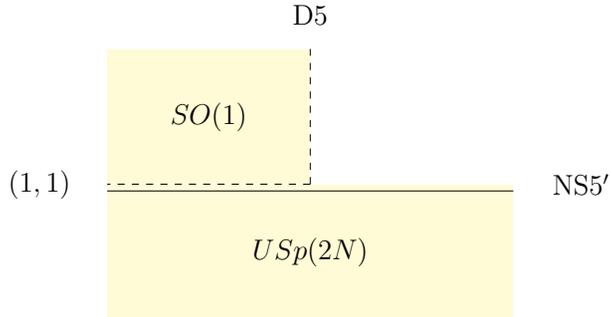

Let us study the Y-junction realizing the ortho-symplectic Y-algebra $Y_{N,0,0}^-$. 
In Type IIB string theory, $N$ D3-branes extend in the region $x^2\le 0$ in the presence of an O3$^+$-plane. 
They terminate on two distinct 5-branes: 
a half NS5-brane at $x^2=0$, $x^6\ge 0$, 
and a half $(1,1)$ 5-brane along $x^2=x^6\le 0$. 
The configuration preserves $\mathcal{N}=(0,4)$ supersymmetry at the junction $x^2=x^6=0$. 
From the field theory perspective, the system contains 4d $\mathcal{N}=4$ SYM theory with gauge group $USp(2N)$ 
supported on $x^2\le 0$, with the Neumann boundary condition $\mathcal{N}'$ imposed at $x^2=0$. 
Although no D3-branes are suspended between the half D5- and $(1,1)$ 5-branes, 
this region is effectively described by a formal $SO(1)$ gauge theory. 
While this gauge factor is trivial and supports no dynamical gauge fields, it encodes the presence of matter degrees of freedom at the interface. 
More precisely, one obtains a half twisted hypermultiplet transforming in the fundamental representation of the adjacent $USp(2N)$ gauge group. 
It obeys the Dirichlet boundary condition, as required by the D5-brane located at $x^2\ge 0$, $x^6=0$. 
We also note that the interaction with the $(1,1)$ 5-brane produces a localized gauge anomaly at the intersection $x^2=x^6=0$, 
which is now compensated by the 3d fundamental twisted hypermultiplet with Dirichlet boundary condition. 
We refer to this configuration as the $\left(
\begin{smallmatrix}
&SO(1)&|&0& \\ \hline 
&&USp(2N)&& \\
\end{smallmatrix}
\right)$ Y-junction. 
The gauge theory configuration is depicted in Figure \ref{fig_Y-1}. 

%
%
%
%
%

The quarter-index for the $\left(
\begin{smallmatrix}
&SO(1)&|&0& \\ \hline 
&&USp(2N)&& \\
\end{smallmatrix}
\right)$ Y-junction is evaluated by the matrix integral of the form
\begin{align}
\label{qind_Y-1}
&
\mathbb{IV}_{\mathcal{N}'\mathcal{D}}
^{\left(
\begin{smallmatrix}
&SO(1)&|&0& \\ \hline 
&&USp(2N)&& \\
\end{smallmatrix}
\right)}(t;q)
\nonumber\\
&=
\frac{1}{2^N N!}\frac{(q)_{\infty}^N}{(q^{\frac12}t^{2};q)_{\infty}^N}
\oint \prod_{i=1}^N \frac{ds_i}{2\pi is_i}\frac{(s_i^{\pm2};q)_{\infty}}{(q^{\frac12}t^{2}s_i^{\pm2};q)_{\infty}}
\prod_{i<j}
\frac{ (s_i^{\pm}s_j^{\pm};q)_{\infty}}
{(q^{\frac12}t^{2}s_i^{\pm}s_j^{\pm};q)_{\infty}}
\prod_{i=1}^N (q^{\frac34}ts_i^{\pm};q)_{\infty}. 
\end{align}

\subsection{$Y_{0,N,0}^-$}
\begin{figure}
\usetikzlibrary{shapes}
\centering
\scalebox{0.9}{
\begin{tikzpicture}
\filldraw [fill=yellow!20!white,draw=white] (-3,2) -- (0,2) -- (0,4) -- (-3,4);
\draw (0,2)[dashed] -- (0,4); 
\draw (0,2)[dashed] -- (-3,2); 
\draw (-3,1.9) -- (3,1.9); 
\node at (0,1) {};
\node at (-1.5,3) {$SO(2N+1)$};
\node at (1.5,3) {};
\node at (4,2) {NS5$'$};
\node at (0,4.5) {D5};
\node at (-4,2) {$(1,1)$};
\end{tikzpicture}
}
\caption{The Y-junction for $Y_{0,N,0}^-$. }
\label{fig_Y-2}
\end{figure}
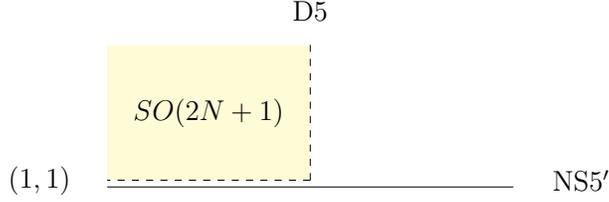

We investigate the Y-junction associated with the Y-algebra $Y_{0,N,0}^-$. 
In Type IIB brane setup, $N$ D3-branes are suspended between 
a half D5-brane and a half $(1,1)$ 5-brane in the presence of an $\widetilde{\textrm{O3}}^-$-plane. 
The field theoretic configuration is realized 
by $\mathcal{N}=4$ $SO(2N+1)$ gauge theory supported on $x^2\ge 0$, subject to the Neumann boundary condition $\mathcal{N}'$ at $x^2=0$. 
At $x^6=0$, the system ends on a singular Nahm pole, which fully breaks the gauge group. 
It is depicted in Figure \ref{fig_Y-2}. 

%
%
%
%
%

We again analyze the quarter-index of this corner configuration using a Higgsing procedure. 
The configuration can be obtained as a deformation of the Neumann-Dirichlet corner for the $SO(2N+1)$ gauge theory, 
and its quarter-index takes the form
\begin{align}
\label{qind_SO2N+1}
&
\mathbb{IV}_{\mathcal{N}' \mathcal{D}}^{SO(2N+1)}(t,x_i;q)
=\frac{1}{(q^{\frac12}t^2;q)_{\infty}^N}
\prod_{i=1}^N \frac{1}{(q^{\frac12}t^2x_i^{\pm};q)_{\infty}}
\prod_{i<j}
\frac{1}{(q^{\frac12}t^2x_i^{\pm}x_j^{\pm};q)_{\infty} }. 
\end{align}
In the present setup, the global fugacities are chosen as follows: 
\begin{align}
x_i&=q^{\frac{i}{2}}t^{2i}. 
\end{align}
We then find that 
\begin{align}
&
\mathbb{IV}_{\mathcal{N}' \mathcal{D}}^{SO(2N+1)}(t,x_i=q^{\frac{i}{2}}t^{2i};q)
\nonumber\\
&=\mathbb{IV}_{\mathcal{N}\mathcal{D}}
^{\left(
\begin{smallmatrix}
&SO(2N+1)&|&0& \\ \hline 
&&0&& \\
\end{smallmatrix}
\right)}(t;q)
\prod_{i=1}^N 
\mathbb{II}_N^{\textrm{3d HM}}(x=q^{\frac{2i-1}{4}}t^{2i-1})
\nonumber\\
&\times 
\prod_{i=1}^{N-1}
\mathbb{II}_{N}^{\textrm{3d HM}}(x=q^{\frac{2i-1}{4}}t^{2i-1})^{N-i}
\prod_{i=1}^{2N-3}
\mathbb{II}_{N}^{\textrm{3d HM}}(x=q^{\frac{2i+3}{4}}t^{2i+3})^{a_N(i)}. 
\end{align}
By decoupling the Neumann half-index of the 3d hypermultiplet, 
we obtain the quarter-index for the $Y_{0,N,0}^-$ junction of the form 
\begin{align}
\label{qind_Y-2}
\mathbb{IV}_{\mathcal{N}\mathcal{D}}
^{\left(
\begin{smallmatrix}
&SO(2N+1)&|&0& \\ \hline 
&&0&& \\
\end{smallmatrix}
\right)}(t;q)
&=
\prod_{k=1}^N \frac{1}{(q^k t^{4k};q)_{\infty}}. 
\end{align}
This Y-junction is related to the $\left(
\begin{smallmatrix}
&SO(1)&|&0& \\ \hline 
&&USp(2N)&& \\
\end{smallmatrix}
\right)$ Y-junction by the $SL(2,\mathbb{Z})$ transformation in Type IIB string theory, 
and is thus expected to be dual as a field theoretic configuration. 
Indeed, the quarter-index (\ref{qind_Y-2}) obtained via the Higgsing precisely matches the matrix integral expression (\ref{qind_Y-1}) of the $\left(
\begin{smallmatrix}
&SO(1)&|&0& \\ \hline 
&&USp(2N)&& \\
\end{smallmatrix}
\right)$ Y-junction.,
\begin{align}
\label{eq:Y-1=Y-2}
\mathbb{IV}_{\mathcal{N}'\mathcal{D}}
^{\left(
\begin{smallmatrix}
&SO(1)&|&0& \\ \hline 
&&USp(2N)&& \\
\end{smallmatrix}
\right)}(t;q)
=
\mathbb{IV}_{\mathcal{N}\mathcal{D}}
^{\left(
\begin{smallmatrix}
&SO(2N+1)&|&0& \\ \hline 
&&0&& \\
\end{smallmatrix}
\right)}(t;q)
\end{align}
This equality can be proved by the Gustafson integral formula \eqref{Gustafson_int}.
We consider the specialization
\begin{equation}
\begin{aligned}
(a,b,c,d)=(q^{\frac{1}{4}} t, -q^{\frac{1}{4}} t, -q^{\frac{3}{4}} t,0),\qquad \sft=q^{\frac{1}{2}}t^2.
\end{aligned}
\end{equation}
Then the integrand in \eqref{Gustafson_int} becomes
\begin{align}
&\prod_{i=1}^N \frac{(s_i^{\pm 2};\sfq)_\infty}
{(q^{\frac{1}{4}} t s_i^{\pm};\sfq)_\infty\,(-q^{\frac{1}{4}} t s_i^{\pm};\sfq)_\infty\,
(-q^{\frac{3}{4}} ts_i^{\pm};\sfq)_\infty}
\prod_{1\le i<j\le N}\frac{(s_i^{\pm}s_j^{\pm};\sfq)_\infty}
{(q^{\frac{1}{2}}t^2 s_i^{\pm}s_j^{\pm};\sfq)_\infty}\notag \\
&=\prod_{i=1}^N  \frac{(s_i^{\pm2};q)_{\infty}(q^{\frac34}ts_i^{\pm};q)_{\infty}}{(q^{\frac12}t^{2}s_i^{\pm2};q)_{\infty}}
 \prod_{1\le i<j\le N}\frac{(s_i^{\pm}s_j^{\pm};\sfq)_\infty}
{(q^{\frac{1}{2}}t^2 s_i^{\pm}s_j^{\pm};\sfq)_\infty},
\label{eq:G-integrand-3}
\end{align}
where we have used the identity \eqref{eq:product-formula-1}.
This is the integrand of the quarter-index \eqref{qind_Y-1}.
The Gustafson integral formula \eqref{Gustafson_int} is now given by
\begin{align}
&G_N(q^{\frac{1}{4}} t, -q^{\frac{1}{4}} t, -q^{\frac{3}{4}} t,0;q, q^{\frac{1}{2}}t^2)\notag \\
&=\frac{1}{2^N N!}
\oint\prod_{i=1}^N \frac{ds_i}{2\pi i s_i}\frac{(s_i^{\pm2};q)_{\infty}(q^{\frac34}ts_i^{\pm};q)_{\infty}}{(q^{\frac12}t^{2}s_i^{\pm2};q)_{\infty}}
\prod_{1\le i<j\le N}\frac{(s_i^{\pm}s_j^{\pm};\sfq)_\infty}
{(q^{\frac{1}{2}}t^2 s_i^{\pm}s_j^{\pm};\sfq)_\infty} \notag\\
&=\frac{(q^{\frac{1}{2}}t^2;q)_\infty^N}{(q;q)_\infty^N}\prod_{j=1}^N
\frac{1}{(q^{\frac{j}{2}}t^{2j};q)_\infty(-q^{\frac{j}{2}}t^{2j};q)_\infty(q^{\frac{j+1}{2}}t^{2j};q)_\infty(-q^{\frac{j+1}{2}}t^{2j};q)_\infty}.
\end{align}
After some computations, one finds that the last product reduces to the simpler one
\begin{align}
\prod_{j=1}^N
\frac{1}{(q^{\frac{j}{2}}t^{2j};q)_\infty(-q^{\frac{j}{2}}t^{2j};q)_\infty(q^{\frac{j+1}{2}}t^{2j};q)_\infty(-q^{\frac{j+1}{2}}t^{2j};q)_\infty}=
\prod_{k=1}^{N} \frac{1}{(q^{k} t^{4k} ;q )_\infty}.
\end{align}
Therefore we finally obtain the equality \eqref{eq:Y-1=Y-2}.

Upon taking the H-twist limit, the quarter-index (\ref{qind_Y-2}) or equivalently (\ref{qind_Y-1}) can be written as
\begin{align}
\label{ch_WB}
\mathbb{IV}_{\mathcal{N}\mathcal{D}}
^{\left(
\begin{smallmatrix}
&SO(2N+1)&|&0& \\ \hline 
&&0&& \\
\end{smallmatrix}
\right)}(t=q^{\frac14};q)
=\prod_{k=1}^{N}\frac{1}{(q^{2k};q)_{\infty}}
=\chi_{\mathcal{W}_{\mathfrak{so}(2N+1)}}. 
\end{align}
This is identified with the vacuum character of the W-algebra $\mathcal{W}_{\mathfrak{so}(2N+1)}$ of type B \cite{Frenkel:1994em,MR1174415}, 
which is realized as the Y-algebra $Y_{0,N,0}^-$ \cite{Gaiotto:2017euk}.

\section{$\tilde{Y}_{N,0,0}^-$ and $Y_{0,N,0}^+$}
\label{sec_C2}

\subsection{$\tilde{Y}_{N,0,0}^-$}
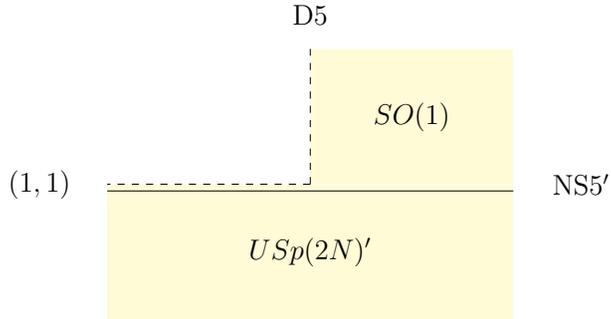
\begin{figure}
\usetikzlibrary{shapes}
\centering
\scalebox{0.9}{
\begin{tikzpicture}
\filldraw [fill=yellow!20!white,draw=white] (-3,0) -- (3,0) -- (3,2) -- (-3,2);
\filldraw [fill=yellow!20!white,draw=white] (0,2) -- (3,2) -- (3,4) -- (0,4);
\draw (0,2)[dashed] -- (0,4); 
\draw (0,2)[dashed] -- (-3,2); 
\draw (-3,1.9) -- (3,1.9); 
\node at (0,1) {$USp(2N)'$};
\node at (-1.5,3) {};
\node at (1.5,3) {$SO(1)$};
\node at (4,2) {NS5$'$};
\node at (0,4.5) {D5};
\node at (-4,2) {$(1,1)$};
\end{tikzpicture}
}
\caption{The Y-junction for $\tilde{Y}_{N,0,0}^-$. }
\label{fig_tY-2}
\end{figure}

Let us consider the Y-junction that can realize the Y-algebra $\tilde{Y}_{N,0,0}^-$. 
In this configuration a stack of $N$ D3-branes in the presence of a $\widetilde{\textrm{O3}}^+$-plane ends on a half NS5-brane at $x^2=0$, $x^6\ge0$. 
In the field theoretic description, this setup realizes $\mathcal{N}=4$ SYM theory with gauge group $USp(2N)$ 
with a non-trivial discrete theta angle, which is denoted by $USp(2N)'$ gauge theory, 
defined on the half-space $x^2\le 0$ with the Neumann boundary condition $\mathcal{N}'$ at $x^2=0$. 
The quadrant region $x^2 \ge 0$, $x^6 \ge 0$ is formally associated with an $SO(1)$ gauge theory, which is trivial. 
Nevertheless, from the field theoretic viewpoint, it gives rise to a half twisted hypermultiplet transforming in the fundamental representation of the neighboring $USp(2N)$ gauge group. 
Due to the presence of the D5-brane at $x^6=0$, the fundamental half twisted hyper should satisfy the Dirichlet boundary condition. 
The corresponding gauge theory configuration is depicted in Figure \ref{fig_tY-2}. 
This gauge theory configuration is related in a manner analogous to the distinction between $\mathcal{N}=4$ $USp(2N)$ and $USp(2N)'$ SYM theories, 
in that they differ in their global data but yield identical contributions to the index.

%
%
%
%
%

Accordingly, the quarter-index takes the form
\begin{align}
\label{qind_tY-2}
&
\mathbb{IV}_{\mathcal{N}'\mathcal{D}}
^{\left(
\begin{smallmatrix}
&0&|&SO(1)& \\ \hline 
&&USp(2N)'&& \\
\end{smallmatrix}
\right)}(t;q)
\nonumber\\
&=
\frac{1}{2^N N!}\frac{(q)_{\infty}^N}{(q^{\frac12}t^{2};q)_{\infty}^N}
\oint \prod_{i=1}^N \frac{ds_i}{2\pi is_i}\frac{(s_i^{\pm2};q)_{\infty}}{(q^{\frac12}t^{2}s_i^{\pm2};q)_{\infty}}
\prod_{i<j}
\frac{ (s_i^{\pm}s_j^{\pm};q)_{\infty}}
{(q^{\frac12}t^{2}s_i^{\pm}s_j^{\pm};q)_{\infty}}
\prod_{i=1}^N (q^{\frac34}ts_i^{\pm};q)_{\infty}. 
\end{align}

\subsection{$Y_{0,N,0}^+$}
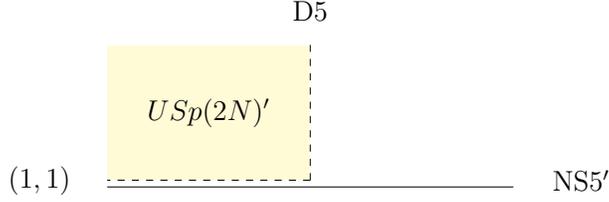
\begin{figure}
\usetikzlibrary{shapes}
\centering
\scalebox{0.9}{
\begin{tikzpicture}
\filldraw [fill=yellow!20!white,draw=white] (-3,2) -- (0,2) -- (0,4) -- (-3,4);
\draw (0,2)[dashed] -- (0,4); 
\draw (0,2)[dashed] -- (-3,2); 
\draw (-3,1.9) -- (3,1.9); 
\node at (0,1) {};
\node at (-1.5,3) {$USp(2N)'$};
\node at (1.5,3) {};
\node at (4,2) {NS5$'$};
\node at (0,4.5) {D5};
\node at (-4,2) {$(1,1)$};
\end{tikzpicture}
}
\caption{The Y-junction for $Y_{0,N,0}^+$. }
\label{fig_Y+2}
\end{figure}

Considering the S-dual of the brane configuration, 
one is led to the Y-junction that is expected to realize the $Y^{+}_{0,N,0}$ algebra. 
In this dual setup, a stack of $N$ D3-branes occupies the region between a half D5-brane and a half $(1,1)$ 5-brane in the presence of an $\widetilde{O3}^+$-plane.
In the absence of the additional 5-branes, this quadrant region supports 4d $\mathcal{N}=4$ SYM $USp(2N)'$ gauge theory. 
However, since the $N$ D3-branes terminate on a single half D5-brane at $x^6=0$, the $USp(2N)$ gauge group is completely broken, 
and the regular Nahm pole boundary condition is imposed.
Furthermore, the presence of a half $(1,1)$ 5-brane imposes the Neumann boundary condition. 
The corresponding gauge theory configuration is shown in Figure \ref{fig_Y+2}.

%
%
%
%
%

We evaluate the quarter-index of this corner configuration by means of the Higgsing procedure. 
The configuration arises as a deformation of the Neumann-Dirichlet corner for the $USp(2N)$ gauge theory\footnote{Strictly speaking, 
the system realizes $USp(2N)'$ gauge theory with a non-trivial discrete theta angle. 
However, in this Y-junction configuration the gauge group is completely broken, and the index is not sensitive to the discrete theta angle, which encodes global (topological) data of the gauge theory. Thus, for the purpose of the index computation, it suffices to treat the system as the Neumann-Dirichlet corner of $USp(2N)$ gauge theory. }, and its quarter-index is given by
\begin{align}
\label{qind_USp2N}
&
\mathbb{IV}_{\mathcal{N}' \mathcal{D}}^{USp(2N)}(t,x_i;q)
=\frac{1}{(q^{\frac12}t^2;q)_{\infty}^N}
\prod_{i=1}^N \frac{1}{(q^{\frac12}t^2x_i^{\pm2};q)_{\infty}}
\prod_{i<j}
\frac{1}{(q^{\frac12}t^2x_i^{\pm}x_j^{\pm};q)_{\infty} }. 
\end{align}
We consider a specialization of the global fugacities such that
\begin{align}
x_i&=q^{\frac{2i-1}{4}}t^{2i-1}. 
\end{align}
It then follows that 
\begin{align}
&
\mathbb{IV}_{\mathcal{N}' \mathcal{D}}^{USp(2N)}(t,x_i=q^{\frac{2i-1}{4}}t^{2i-1};q)
\nonumber\\
&=\mathbb{IV}_{\mathcal{N}\mathcal{D}}
^{\left(
\begin{smallmatrix}
&USp(2N)'&|&0& \\ \hline 
&&0&& \\
\end{smallmatrix}
\right)}(t;q)
\prod_{i=1}^N 
\mathbb{II}_N^{\textrm{3d HM}}(x=q^{\frac{2i-1}{2}}t^{4i-2})
\nonumber\\
&\times 
\prod_{i=1}^{N-1}
\mathbb{II}_{N}^{\textrm{3d HM}}(x=q^{\frac{2i-1}{4}}t^{2i-1})^{N-i}
\prod_{i=1}^{2N-3}
\mathbb{II}_{N}^{\textrm{3d HM}}(x=q^{\frac{2i+1}{4}}t^{2i+1})^{a_N(i)}. 
\end{align}
By stripping off the contributions of the Neumann half-index associated with the 3d hypermultiplet, 
one arrives at the quarter-index corresponding to the Y-junction for $Y_{0,N,0}^{+}$, which takes the following form: 
\begin{align}
\label{qind_Y+2}
\mathbb{IV}_{\mathcal{N}\mathcal{D}}
^{\left(
\begin{smallmatrix}
&USp(2N)'&|&0& \\ \hline 
&&0&& \\
\end{smallmatrix}
\right)}(t;q)
&=\prod_{k=1}^{N}\frac{1}{(q^k t^{4k};q)_{\infty}}. 
\end{align}
The agreement between the two quarter-indices (\ref{qind_tY-2}) and (\ref{qind_Y+2}), as predicted by S-duality, 
is essentially identical to the proof presented in section \ref{sec_C}. 

Upon implementing the H-twist, the quarter-index (\ref{qind_tY-2}) or equivalently (\ref{qind_Y+2}) takes the following form: 
\begin{align}
\label{ch_WC}
\mathbb{IV}_{\mathcal{N}\mathcal{D}}
^{\left(
\begin{smallmatrix}
&USp(2N)'&|&0& \\ \hline 
&&0&& \\
\end{smallmatrix}
\right)}(t=q^{\frac14};q)
&=\prod_{k=1}^N\frac{1}{(q^{2k};q)_{\infty}}=\chi_{\mathcal{W}_{\mathfrak{usp}(2N)}}. 
\end{align}
This expression agrees with the vacuum character of the W-algebra $\mathcal{W}_{\mathfrak{usp}(2N)}$ of type C \cite{Frenkel:1994em,MR1174415}. 
This is consistent with the fact that, for $USp(2N)$ gauge theory, the Neumann-Nahm junction gives rise, via the quantum Drinfeld-Sokolov reduction, 
to the W-algebra of type C as the corner algebra $Y_{0,N,0}^+$ \cite{Gaiotto:2017euk}.
It also coincides in form with that of the W-algebra of type B given by (\ref{ch_WB}). 
The equality of the vacuum characters for the W-algebras of type B and type C follows from the fact that 
the sets of Casimir invariants have identical degrees though the underlying VOAs are not isomorphic.

\acknowledgments{
The work of Y.H. was supported in part by JSPS KAKENHI Grant Nos. 22K03641 and 23K25790.
The work of T.O. was supported by the Startup Funding no.\ 4007012317 of the Southeast University. 
}

\bibliographystyle{utphys}
\bibliography{ref}

\end{document}